\documentclass[10pt]{article} 
\usepackage[preprint]{tmlr}

\usepackage{amsmath,amsfonts,bm}

\def\eqref#1{equation~\ref{#1}}

\def\1{\bm{1}}

\DeclareMathAlphabet{\mathsfit}{\encodingdefault}{\sfdefault}{m}{sl}
\SetMathAlphabet{\mathsfit}{bold}{\encodingdefault}{\sfdefault}{bx}{n}

\DeclareMathOperator*{\argmax}{arg\,max}

\usepackage[utf8]{inputenc}
\usepackage{xspace}

\usepackage{amsmath,amsthm,amssymb}
\usepackage{mathtools}
\usepackage{altvars}

\usepackage[linesnumbered,ruled,vlined]{algorithm2e}
\SetKwInput{KwInput}{Input}                
\SetKwInput{KwOutput}{Output}              

\usepackage[hidelinks]{hyperref}
\usepackage[capitalise]{cleveref}
\crefname{equation}{}{}
\Crefname{equation}{Equation}{Equations}
\crefname{figure}{Fig.}{Figs.}

\usepackage{xcolor}
\definecolor{heteroppca}{RGB}{ 53, 96,240}
\definecolor{homoppca}{RGB}{255,160,253}
\definecolor{group1}  {RGB}{ 73,227,176}
\definecolor{group2}  {RGB}{255,166, 23}
\definecolor{inv}  {RGB}{ 10,112, 56}
\definecolor{sqinv}{RGB}{247,  5,  5}
\definecolor{heteropca}{RGB}{115,222,255}
\definecolor{heteroppcaknown}{RGB}{255,194, 38}

\usepackage[caption=false]{subfig}
\usepackage{graphicx}
\usepackage{tikz,pgfplots}
\pgfplotsset{compat=1.16}

\usepackage[shortlabels]{enumitem}

\usepackage[bb=boondox]{mathalfa}

\newcommand{\revise}[1] {{\color{black}#1}}
\newcommand{\tmlrrev}[1] {{\color{black}#1}}

\newcommand{\tr}{\operatorname{tr}}

\newcommand{\xmath}[1] {\ensuremath{#1}\xspace}
\newcommand{\defequ} {\triangleq}

\long\def\red#1{\bgroup\color{red}#1\egroup}


\newtheorem*{assumption*}{\assumptionnumber}
\providecommand{\assumptionnumber}{}


\newcommand{\Psib} {\bar{\Psi}}

\newcommand{\sumigil} {\sum_{i \, : \, g_i = \ell}}

\usepackage{hyperref}
\usepackage{url}

\title{Streaming Heteroscedastic Probabilistic PCA
\\
with Missing Data}

\author{
\name Kyle Gilman \email kgilman@umich.edu \\
\addr Department of Electrical Engineering and Computer Science \\
University of Michigan
\AND
\name David Hong \\
\addr Department of Electrical and Computer Engineering \\ University of Delaware 
\AND
\name Jeffrey A. Fessler \addr Department of Electrical Engineering and Computer Science \\
University of Michigan
\AND
\name Laura Balzano
\addr Department of Electrical Engineering and Computer Science \\
University of Michigan
}

\def\month{MM}  
\def\year{YYYY} 
\def\openreview{\url{https://openreview.net/forum?id=XXXX}} 

\newcommand{\oneortwocol}[2]{#2}

\begin{document}

\maketitle

\begin{abstract}
Streaming principal component analysis (PCA) is an integral tool
in large-scale machine learning
for rapidly estimating low-dimensional subspaces
from very high-dimensional
data arriving at a high rate.
However, modern datasets increasingly combine data from a variety of sources,
and thus may exhibit heterogeneous quality across samples.
Standard streaming PCA algorithms do not account for non-uniform noise,
so their subspace estimates can quickly degrade.
While the recently proposed
Heteroscedastic Probabilistic PCA Technique (HePPCAT) addresses this heterogeneity,
it was not designed to handle streaming data,
which may exhibit non-stationary behavior.
Moreover, HePPCAT does not allow for missing entries in the data,
which can be common in streaming data.
This paper proposes the Streaming HeteroscedASTic Algorithm for PCA (SHASTA-PCA)
to bridge this divide.
SHASTA-PCA employs a stochastic alternating expectation maximization approach
that jointly learns the low-rank latent factors and the unknown noise variances
from streaming data that may have missing entries and heteroscedastic noise,
all while maintaining a low memory and computational footprint.
Numerical experiments demonstrate the superior subspace estimation
of our method compared to state-of-the-art streaming PCA algorithms 
in the heteroscedastic setting.
Finally, we illustrate SHASTA-PCA applied to highly heterogeneous real data from astronomy. 

\end{abstract}

\section{Introduction}
\label{sec:intro}

Modern data are increasingly large in scale
and formed by combining heterogeneous samples from diverse sources
or conditions that exhibit heteroscedastic noise,
or noises of different variances \citep{hong2021heppcat},
such as in astronomy \citep{ahumada202016th},
medical imaging \citep{pruessmann1999sense, anam2020CTmeanfilter},
and cryo-electron microscopy imaging \citep{anden2017factor, bendory2020single}.
Principal component analysis (PCA) for visualization, exploratory data analysis,
data compression, predictive tasks, or other downstream tasks
is often a fundamental tool to process these high-dimensional data.
\tmlrrev{However, several practical challenges arise when computing PCA on these types of data.}
In many applications,
due to memory or physical constraints,
the full data cannot be observed in their entirety at computation time
and are instead read partially into memory piece by piece,
or observations may stream in continuously and indefinitely.
Moreover, the signal may evolve over time
and require adaptive tracking algorithms for the low-rank component.
Adding to these difficulties,
it is also common for big data to contain missing entries,
such as in magnetic resonance imaging \citep{mensch2017stochastic},
collaborative filtering \citep{candesplan2010matrixcompletion},
and environmental sensing \citep{sensor_network_data_fault_types2009}.
Consequently, there is a need for scalable streaming PCA techniques that can handle
heteroscedastic noise and missing data.

A tremendous body of work has studied streaming PCA techniques
for learning a signal subspace
from noisy incremental data observations that have missing entries.
Streaming or online PCA algorithms often enjoy the advantages of computational efficiency,
low memory overhead, and adaptive tracking abilities,
making them very useful in real-world big-data applications.
However, no existing streaming methods account for noise with differing variances across samples,
i.e., sample-wise heteroscedastic noise,
and their subspace estimates can be highly corrupted by the noisiest samples.
The work in~\citet{hong2021heppcat}
developed a Heteroscedastic Probabilistic PCA technique (HePPCAT)
for data with varying noise levels across the samples.
HePPCAT learns the low-rank factors and the unknown noise variances
via maximum-likelihood estimation,
but only in the batch setting with no missing entries.
Other batch heteroscedastic PCA algorithms,
like weighted PCA studied in \citet{jolliffe2002pca, young1941mle, hong2023owp}
and HeteroPCA \citep{zhang:18:hpa} for data with heteroscedastic features,
also lack streaming and adaptive tracking abilities.
\tmlrrev{None of the existing methods handle all of the real-data complexities we study here: missing entries,
heteroscdastic noise, and streaming data.
Tackling this non-trivial setting requires developing new algorithms.}

To the best of our knowledge,
this paper is the first work to develop a streaming PCA algorithm
for data with missing entries and heteroscedastic noise.
Our algorithm jointly estimates the factors and unknown noise variances
in an \emph{online} fashion from streaming incomplete data
using an alternating stochastic minorize-maximize (SMM) approach
with small computational and memory overhead.
\tmlrrev{We carefully design minorizers
with a particular alternating schedule of stochastic updates
that distinguishes our approach from existing SMM methods.
Notably, handling missing entries and heteroscedastic noise involves more complex updates than the simpler algebraic formulations of algorithms
like HePPCAT or PETRELS \citep{chi_petrels2013}.}
We demonstrate that our algorithm can estimate the signal subspace
from subsampled data
(even without knowing the noise variances)
better than state-of-the-art streaming PCA methods that assume homogeneous noise.
Our algorithm \tmlrrev{is unique in that it} not only tracks low-dimensional dynamic subspaces,
but can also track dynamic noise variances that can occur, e.g.,
in sensor calibration \citep{jun2003drift}
\tmlrrev{and beamforming in nonstationary noise environments \citep{cohen2004nonstationary}}. 
\tmlrrev{Finally, this work extends our understanding of streaming PCA
to the setting of heteroscedastic noise
and draws connections to existing work in the literature. 
In particular, the proposed method closely relates
to the streaming PCA algorithm PETRELS \citep{chi_petrels2013},
but differs by learning the noise variances of the data on-the-fly
and downweighting noisier data samples in the factor updates.
Furthermore, we show that our proposed method
implicitly optimizes a regularized least-squares problem
whose adaptive hyperparameter varies by the learned heterogeneous noise variances.
}

\Cref{shasta:sec:rel} discusses related works for both streaming PCA and heteteroscedastic PCA.
\Cref{sec:sppca,sec:emminor,chap:shasta:alg:batch,sshasta} describe the model we consider, define the resulting optimization problem, and derive the proposed algorithm.
\Cref{sec:results} presents synthetic and real data experiments
that demonstrate the benefits of the proposed method
over existing state-of-the-art streaming PCA algorithms. 

\subsection{Notation}

We use bold upper case letters $\bmA$ to denote matrices,
bold lower case letters $\bmv$ to denote vectors,
and non-bold lower case letters $c$ for scalars.
We denote the Hermitian transpose of a matrix as $\bmA'$ and the trace of a matrix as $\tr(\bmA)$.
The Euclidean norm is denoted by $\|\cdot\|_2$.
The identity matrix of size $d \times d$ is denoted as $\bmI_d$.
The notation $i \in [k]$ means $i \in \{1,\dots,k\}$.
\section{Related work} \label{shasta:sec:rel}

\subsection{Streaming PCA}


A rich body of work has developed and investigated a variety of streaming PCA algorithms
for learning a signal subspace from incremental and possibly incomplete data observations.
These methods, however, assume the data have homogeneous quality
and do not model heteroscedastic properties like those considered in this paper.
Since there are too many related works to detail here
(see, e.g., \citet{balzano2018streaming}, for a recent survey),
we highlight a few of the most related. 

One prominent branch of algorithms use stochastic gradient optimization approaches
to update the learned subspace based on a new data observation at each iteration;
see, e.g., \citet{bertsekas2011incremental} and \citet{bottou2010}.
\citet{mardani2015} use stochastic gradient descent
to learn matrix and tensor factorization models
in the presence of missing data
and also include an exponentially weighted data term
that trades off adapting to new data with fitting historical data.
Stochastic gradient descent over Riemannian manifolds is also a popular approach;
see, e.g., \citet{Bonnabel2013, grouse, grasta, goes2014robust}.
Oja's method \citep{Oja1982SimplifiedNM} takes a stochastic gradient step
to update the subspace basis from the most recent data vector,
obtaining a new orthonormal basis after orthogonalization. 
\tmlrrev{\cite{balzano2022equivalence}
proved the equivalence of the GROUSE algorithm \citep{grouse}
with Oja's method for a certain step size.
Other progress on improving and understanding Oja's method has recently been made,
such as an algorithm to adaptively select the learning rate
in one pass over the data \citep{henriksen2019adaoja}
and an analysis of the convergence rate
for non-i.i.d. data sampled from a Markov chain \citep{kumar2024streaming}.}

Some streaming PCA methods share commonalities with quasi-second-order optimization methods.
For example, the PETRELS algorithm proposed in \citet{chi_petrels2013}
fits a factor model to data with missing entries
via a stochastic quasi-Newton method.
PETRELS has computationally efficient updates
but can encounter numerical instability issues in practice after a large number of samples.

\tmlrrev{More recently, streaming PCA and its analogs have been extended to a variety of new problems.
\cite{giannakis2023learning} propose
a streaming algorithm for forecasting dynamical systems
that they show is a type of streaming PCA problem.
Streaming algorithms have been proposed for robust PCA \citep{diakonikolas2023nearly},
federated learning and differential privacy \citep{grammenos2020federated},
a distributed Krasulina's method \citep{raja2022distributed},
and probabilistic PCA to track nonstationary processes \citep{lu2024streaming}.
Although not strictly a streaming algorithm,
\cite{blocker2023dynamic} also estimate dynamic subspaces
but by using a piecewise-geodesic model on the Grassmann manifold.
}\tmlrrev{All of these approaches implicitly assume that the data quality is homogeneous across the dataset,
in contrast to our proposed approach.}

\subsection{Stochastic MM methods}

Another vein of work on streaming algorithms,
which has the closest similarities to this paper,
is stochastic majorization-minimization (SMM) algorithms for matrix and tensor factorization.
MM methods construct surrogate functions
that are more easily optimized than the original objective;
\tmlrrev{algorithms that successively optimize these surrogates
are guaranteed to converge to stationary points
of the original objective function
under suitable regularity conditions
\citep{jacobson:07:aet,lange2016mm}.
}
SMM algorithms such as
\citet{Mairal2013_stochastic_mm, Strohmeier2020OnlineNT,lyu2024stochastic}
optimize an approximation of a surrogate
computed from accumulated stochastic surrogates
after observing a new data sample at each iteration.
The work in \citet{Mairal2013_stochastic_mm}
proved almost sure convergence to a stationary point
for non-convex objectives with one block of variables for the SMM technique.
The work in \citet{mensch2017stochastic} proposes subsampled online matrix factorization (SOMF)
for large-scale streaming subsampled data
and gives convergence guarantees under mild assumptions.
In \citet{Strohmeier2020OnlineNT,lyu2024stochastic},
the authors extend SMM to functions that are multi-convex
in blocks of variables for online tensor factorization.
Their framework performs block-coordinate minimization
of a single majorizer at each time point.
They prove almost sure convergence of the iterates to a stationary point
assuming the expected loss function is continuously differentiable with a Lipschitz gradient,
the sequence of weights decay at a certain rate,
and the data tensors form a Markov chain with a unique stationary distribution.
\tmlrrev{More recently, \cite{phan2024stochastic} propose and analyze
several stochastic variance-reduced MM algorithms.}
Like these works, our paper also draws upon SMM techniques;
we use an alternating SMM approach to optimize the log-likelihood function for our model.
\tmlrrev{However, our setting and approach differ from these existing methods in key ways; we detail these differences and their impact on the related convergence theory in \S\ref{sss:convergence}.}

A close analogue to SMM
is the Doubly Stochastic Successive Convex (DSSC) approximation algorithm \citep{mokhtari_koppel2020:dssc}
that optimizes convex surrogates to non-convex objective functions
from streaming samples or minibatches.
A key feature of their algorithm is that it decomposes the optimization variable into $B$ blocks
and operates on random subsets of blocks at each iteration.
Specifically, the DSSC algorithm chooses a block $i \in [B]$,
computes stochastic gradients with respect to the $i \mathrm{th}$ block of variables
and then recursively updates the approximation to the $i\mathrm{th}$ surrogate function.
From the optimizer to the approximate surrogate,
their algorithm performs momentum updates of the iterates very similarly to SMM.
Our own algorithm SHASTA-PCA in \S\ref{s,shasta} can be interpreted as following a similar approach, \tmlrrev{but without using gradient methods since our problem does not have Lipschitz-continuous gradients.}


\subsection{Heterogeneous data}

Several recently proposed PCA algorithms consider data
contaminated by heteroscedastic noise \textit{across samples},
which is the setting we study in this paper.
Weighted PCA is a natural approach in this context \citep{jolliffe2002pca},
either weighting the samples by the inverse noise variances \citep{young1941mle}
or by an optimal weighting derived in \citet{hong2023owp}.
In both instances, the variances must be known \textit{a priori} or estimated to compute the weights.
Probabilistic PCA (PPCA) \citep{tipping1999ppc}
uses a probabilistic interpretation of PCA via a factor analysis model
with isotropic Gaussian noise and latent variables.
For a single unknown noise variance (i.e., homoscedastic noise),
the learned factors and noise variance are solutions to a maximum-likelihood problem
that can be optimized using an expectation maximization algorithm;
these solutions correspond exactly to PCA.
\citet{hong2021heppcat} studied the heteroscedastic probabilistic PCA problem
that considers a factor model where groups of data may have different (unknown) noise variances.
Their method, HePPCAT, performs maximum-likelihood estimation
of the latent factors and unknown noise variances
(assuming knowledge of which samples belong to each noise variance group);
they consider various algorithms and recommend an alternating EM approach. 
Other batch heteroscedastic PCA methods have since followed HePPCAT.
ALPACAH \citep{cavazos2023alpcah} estimates the low-rank component and variances
for data with sample-wise heteroscedastic noise,
but because their objective function is not separable by the samples,
no streaming counterpart currently exists.
HeMPPCAT \citep{xu_balzano_fessler23:hemppcat}
extends mixtures of probabilistic PCA \citep{tipping1999mixtures}
to the case of heteroscedastic noise across samples.

More broadly, there is an increasing body of work
that investigates PCA techniques for data contaminated by some sort of heterogeneous noise,
including noise that is heteroscedastic \textit{across features}.
HeteroPCA \citep{zhang:18:hpa} iteratively imputes the diagonal entries of the sample covariance matrix
to address the bias in these entries that arises
when the noise has feature-wise heteroscedasticity,
\tmlrrev{and \cite{zhou2023deflated} then extended HeteroPCA to the case of ill-conditioned low-rank data.}
Another line of work in \citet{leeb2021optimal} and \citet{leeb2021matrix}
has considered rescaling the data to instead whiten the noise.
\tmlrrev{\cite{yan2024inference} develops inference and uncertainty quantification procedures
for PCA with missing data and feature-wise heteroscedasticity.}
There has also been recent progress on developing methods
to estimate the rank in heterogeneous noise contexts \citep{hong2020stn:arxiv:v1,ke2021eot,landa2022brt,landa2025dyson}
and on establishing fundamental limits for recovery in these settings \citep{behne2022fundamental,zhang2024matrix}.

A closely related problem in signal processing applications,
such as heterogeneous clutter in radar,
is data with heterogeneous ``textures'', also called the ``mixed effects'' problem.
Here, the signal is modeled as a mixture of scaled Gaussians,
each sharing a common low-rank covariance scaled
by an unknown deterministic positive ``texture'' or power factor \citep{breloy2019icb}.
In fact, the heterogeneous texture and HPPCA problems
are related up to an unknown scaling \citep{hong2021heppcat}. \citet{ferrer2021robust}
and \citet{HIPPERTFERRER2022108460} also studied variations of the heterogeneous texture problem 
for robust covariance matrix estimation from batch data with missing entries.
\citet{hppca_collas2021} study the probabilistic PCA problem
in the context of isotropic signals with unknown heterogeneous textures and a known noise floor.
Their paper casts the maximum-likelihood estimation as an optimization problem over a Riemannian manifold,
using gradient descent on the manifold to jointly optimize for the subspace and the textures.
Their formulation also readily admits a stochastic gradient algorithm for online optimization. 

Heteroscedastic data has also been investigated in the setting of supervised learning
for fitting linear regression models with stochastic gradient descent \citep{Song2015}.
The authors show that the model's performance given ``clean'' and ``noisy'' datasets
depends on the learning rate
and the order in which the datasets are processed.
Further, they propose using separate learning rates
that depend on the noise levels
instead of using one learning rate as is done in classical SGD.
\tmlrrev{In the context of generalized linear bandits,
\cite{zhao2023optimal} propose an online algorithm for the heteroscedastic bandit problem
using weighted linear regression with weights selected as the inverse noise variance.}

\section{Probabilistic Model} \label{sec:sppca}

Similar to
\citet{hong2019ppf,hong2021heppcat},
we model data samples in $\bbR^d$
from $L$ noise level groups as:
\begin{equation}
  \bmy_{i} = \bmF \bmz_{i} + \bmvarepsilon_{i}
  ,
  \qquad
  \text{for }
  i = 1,2,\ldots
  ,
   \label{eq:shasta:generative_model}
\end{equation}
where
$\bmF \in \bbR^{d \times k}$ is a deterministic factor matrix to estimate,
$\bmz_{i} \sim \clN(\bm0_k,\bmI_k)$
are independent and identically distributed (i.i.d.) coefficient vectors,
$\bmvarepsilon_{i} \sim \clN(\bm0_d,v_{g_i} \bmI_d)$ are i.i.d. noise vectors,
$g_i \in \{1,\dots,L\}$ is the noise level group to which the $i$th sample belongs,
and $v_1,\dots,v_L$ are deterministic noise variances to estimate.
We assume the group memberships $g_i$ are known.

Let $\Omega_i \subseteq \{1,\hdots,d\}$
denote the set of entries observed for the $i$th sample,
and let $\bmy_{\Omega_i} \in \bbR^{|\Omega_i|}$
and $\bmF_{\Omega_i} \in \bbR^{|\Omega_i| \times k}$
denote the restrictions of $\bmy_i$ and $\bmF$ to the entries and rows defined by $\Omega_i$.
Then the observed entries of the data vectors are distributed as
\begin{equation*}
  \bmy_{\Omega_i} \sim \clN(\bm0_{|\Omega_i|}, \bmF_{\Omega_i} \bmF_{\Omega_i}' + v_{g_i} \bmI_{|\Omega_i|}).
\end{equation*}

We will express the joint log-likelihood over only the \textit{observed} entries of the data
and maximize it for the unknown deterministic model parameters.

For a batch of $n$ vectors,
the joint log-likelihood over the observed batch data
for $\Omega = (\Omega_1,\hdots,\Omega_n)$
can be easily written in an incremental form as a sum of log-likelihoods
over the partially observed dataset
$\bmY_\Omega \defequ (\bmy_{\Omega_1}, \dots, \bmy_{\Omega_n})$:
\begin{align}
\label{eq:log_likelihood:incremental}
    \clL(\bmY_\Omega; \bmF,\bmv)
   &= \frac{1}{2} \sum_{i=1}^n \clL_i(\bmy_{\Omega_i}; \bmF, \bmv) + C,
\end{align}
where $C$ is a constant independent of $\bmF$ and $\bmv$, and
\begin{align}
    \clL_i(\bmy_{\Omega_i}; \bmF, \bmv)
    \defequ
    \ln\det(\bmF_{\Omega_i}\bmF_{\Omega_i}' + v_{g_i} \bmI_{|\Omega_i|})^{-1}
    -
    \tr\big\{
        \bmy_{\Omega_i}'(\bmF_{\Omega_i}\bmF_{\Omega_i}'
        + v_{g_i} \bmI_{|\Omega_i|})^{-1}\bmy_{\Omega_i}
    \big\}
    ,
\end{align}
is the loss for a single vector $\bmy_{\Omega_i}$. To jointly estimate the factor matrix $\bmF$
and the variances $\bmv$,
we maximize this likelihood. 
Optimizing the log-likelihood \cref{eq:log_likelihood:incremental}
is a challenging non-concave optimization problem,
so we propose an efficient alternating minorize-maximize (MM) approach.

\section{Expectation Maximization Minorizer}
\label{sec:emminor}


This section derives a minorizer for the log-likelihood \cref{eq:log_likelihood:incremental}
that will be used to develop the proposed alternating MM algorithm in the following sections.
In particular, we derive a minorizer
at the point $(\btlF,\btlv)$
in the style of expectation maximization methods.
The minorizer follows from the work in \citet{hong2021heppcat};
here we extend it to the case for data with missing entries.

For the complete-data log-likelihood,
we use the observed samples $\bmy_{\Omega_i}$ and unknown coefficients $\bmz_i$,
leading to the following complete-data log-likelihood for the $i$th sample:
\begin{align}
  \clL_i^c(\bmF,\bmv)
    &
  \defequ
  \ln p(\bmy_{\Omega_i}, \bmz_i; \bmF, \bmv)
  \nonumber \\ &
  = \ln p(\bmy_{\Omega_i} | \bmz_i; \bmF, \bmv) + \ln p(\bmz_i; \bmF, \bmv)
  \nonumber \\
  &
  =
    - \frac{|\Omega_i|}{2} \ln v_{g_i}
    - \frac{\|\bmy_{\Omega_i} - \bmF_{\Omega_i}\bmz_i\|_2^2}{2v_{g_i}}
    - \frac{\|\bmz_i\|_2^2}{2}
  , \label{eq:em:complete_likelihood_i}
\end{align}
where \cref{eq:em:complete_likelihood_i} drops
the constants $\ln(2\pi)^{-|\Omega_i|/2}$ and $\ln(2\pi)^{-k/2}$.

Next, we take the expectation of \cref{eq:em:complete_likelihood_i}
with respect to the following conditionally independent distributions
obtained from Bayes' rule and the matrix inversion lemma:
\begin{equation} \label{eq:shasta:em:conddist}
  \bmz_i | \{ \bmy_{\Omega_i}, \bmF = \btlF, \bmv = \btlv \}
  \overset{\text{ind}}{\sim}
  \clN(
    \bmM_i(\btlF,\btlv) \btlF_{\Omega_i}' \bmy_{\Omega_i}, \tlv_{g_i} \bmM_{i}(\btlF,\btlv)
  )
  ,
\end{equation}
where
$\bmM_i(\btlF,\btlv) \defequ (\btlF_{\Omega_i}'\btlF_{\Omega_i} + \tlv_{g_i}\bmI_k)^{-1}$. 
%
Doing so yields the following minorizer for $\clL_i$
at $(\btlF,\btlv)$:
\begin{equation} 
\label{eq:shasta:em:estep}
\begin{split}
  &
  \Psi_i(\bmF, \bmv; \btlF, \btlv)
  \defequ 
      - \frac{|\Omega_i|}{2} \ln v_{g_i}
      - \frac{\|\bmy_{\Omega_i}\|_2^2}{2 v_{g_i}}
      + \frac{1}{v_{g_i}} \bmy_{\Omega_i}' \bmF_{\Omega_i} \bbrz_{i}(\btlF,\btlv)
      \oneortwocol{\\&\qquad \qquad \qquad}{\\&\qquad \qquad \qquad}
      - \frac{1}{2 v_{g_i}}\left(
        \|\bmF_{\Omega_i}
        \bbrz_{i}(\btlF,\btlv)\|^2_2 + \tlv_{g_i} \tr\{\bmF_{\Omega_i}'\bmF_{\Omega_i}\bmM_{i}(\btlF,\btlv)\}\right)
  ,
  \end{split}
\end{equation}
where
$\bbrz_{i}(\btlF,\btlv)
\defequ \bmM_{i}(\btlF,\btlv) \btlF_{\Omega_i}' \bmy_{\Omega_i}$
and
\cref{eq:shasta:em:estep} drops terms
that are constant with respect to $\bmF$ and $\bmv$.

\section{A Batch Algorithm}
\label{chap:shasta:alg:batch}

Before deriving the proposed streaming algorithm, SHASTA-PCA,
we first derive a batch method for comparison purposes.
Summing the sample-wise minorizer \cref{eq:shasta:em:estep}
across all the samples
gives the following batch minorizer
at the point $(\btlF,\btlv)$:
\begin{align} \label{eq:em:batch_alg}
  \Psi(\bmF, \bmv; \btlF, \btlv)
  &\defequ \sum_{i=1}^n \Psi_i(\bmF,\bmv; \btlF,\btlv)\\
  &\nonumber = 
      \sum_{\ell=1}^L \sumigil - \frac{|\Omega_{i}|}{2} \ln v_{\ell}
      - \frac{\|\bmy_{\Omega_{i}}\|_2^2}{2 v_{\ell}}
      + \frac{1}{v_{\ell}} \bmy_{\Omega_{i}}' \bmF_{\Omega_{i}} \bbrz_{i}(\btlF,\btlv)
      \oneortwocol{}{\\&\quad}
      - \frac{1}{2 v_{\ell}}\left(
        \|\bmF_{\Omega_{i}}
        \bbrz_{i}(\btlF,\btlv)\|^2_2 + \tlv_{\ell}
        \tr\{\bmF_{\Omega_{i}}'\bmF_{\Omega_{i}}\bmM_{i}(\btlF,\btlv)\}\right) \nonumber
  .
\end{align}
Similar to HePPCAT \citep{hong2021heppcat}, which is a batch method for fully sampled data,
in each iteration $t$,
we first
update $\bmv$ (with $\bmF$ fixed at $\bmF_{t-1}$)
then
update $\bmF$ (with $\bmv$ fixed at $\bmv_{t}$),
i.e.,
\begin{align}
    \bmv_{t} &= \argmax_{\bmv} \Psi(\bmF_{t-1}, \bmv; \bmF_{t-1}, \bmv_{t-1})
    , \label{eq:batch:mm:alternative:v}\\
    \bmF_{t} &= \argmax_{\bmF} \Psi(\bmF, \bmv_{t}; \bmF_{t-1}, \bmv_{t})
    . \label{eq:batch:mm:alternative:F}
\end{align}
Here $t$ denotes only the algorithm iteration,
in contrast to the streaming algorithm
in \S\ref{s,shasta},
where $t$ denotes both the time index and algorithm iteration.
The following subsections
derive efficient formulas for these updates
and discuss the memory and computational costs.

\subsection{Optimizing \texorpdfstring{$\bmv$}{v}
for fixed \texorpdfstring{$\bmF$}{F}}

Here we derive an efficient formula
for the $\bmv$ update in \cref{eq:batch:mm:alternative:v}.
While the update is similar to HePPCAT \citep{hong2021heppcat},
the key difference lies in computing the minorizer parameters.
Specifically, the missing data introduces the sample-wise quantities $\btlz_{i}$ and $\btlM_{i}$ below,
which depend on the sampling patterns for the $i\mathrm{th}$ data vector
and must be computed for every sample $i \in [n]$ per iteration
compared to the single $\btlz$ and $\btlM$ used in HePPCAT.
This update
separates into $L$ univariate optimization problems, one in each variance $v_\ell$:
\begin{equation}
    v_{t,\ell}
    =
    \argmax_{v_\ell}
    - \frac{\theta_\ell}{2} \ln v_{\ell}
    - \frac{\tlrho_\ell}{2 v_{\ell}}
    ,
\end{equation}
where
\begin{equation}
 \begin{aligned}
     \theta_\ell &\triangleq \sumigil |\Omega_{i}|, &
     \tlrho_{\ell}
     &\triangleq
    \sumigil \bigg[
        \|\bmy_{\Omega_{i}} - \bmF_{t-1,\Omega_{i}} \btlz_{i}\|_2^2
        +
        v_{t-1,\ell} \tr(\bmF_{t-1,\Omega_{i}}'\bmF_{t-1,\Omega_{i}} \btlM_{i})
    \bigg]
    ,
 \end{aligned}
\end{equation}
$\bmF_{t-1,\Omega_{i}}$ denotes the iterate $\bmF_{t-1}$
restricted to the rows defined by $\Omega_i$,
and we use the following shorthand in this section
\begin{equation}
\begin{aligned}
    \label{eq:btlz_btlM:v_update:batch}
    \btlz_{i} &\defequ \bbrz_{i}(\bmF_{t-1},\bmv_{t-1})
    , &
    \btlM_i &\defequ \bmM_i(\bmF_{t-1},\bmv_{t-1})
    .
\end{aligned}
\end{equation}
The corresponding solutions are
\begin{align}
    v_{t,\ell}
    = \frac{\tlrho_{\ell}}{\theta_\ell}
    .
\end{align}
We precompute
$\theta_\ell$
because it remains constant across iterations.

\subsection{Optimizing \texorpdfstring{$\bmF$}{F}
  for fixed \texorpdfstring{$\bmv$}{v}}

Here we derive an efficient formula
for the $\bmF$ update in \cref{eq:batch:mm:alternative:F}.
The update differs from the factor update in HePPCAT again due to the missing entries in the data.
Specifically, this update
separates into $d$
quadratic optimization problems,
one in each row $\bmf_j$ of $\bmF$:
\begin{align}
    \bmf_{t,j} = \argmax_{\bmf_j} \; \bmf_j' \btls_j - \frac{1}{2} \bmf_j' \btlR_j \bmf_j
    ,
\end{align}
where
\begin{align}
    \btlR_j &\defequ \sum_{\ell=1}^L
    \sum_{\substack{i \, : \, g_i = \ell \\ \phantom{i \, : \,} \Omega_i \ni j}}
    \frac{1}{v_{t,\ell}} (\btlz_{i} \btlz_{i}' + v_{t,\ell} \btlM_{i})\\
    \btls_j &\defequ \sum_{\ell=1}^L
    \sum_{\substack{i \, : \, g_i = \ell \\ \phantom{i \, : \,} \Omega_i \ni j}}
    \frac{1}{v_{t,\ell}} y_{ij} \btlz_{i},
\end{align}
$y_{ij}$ is the $j$th coordinate of the vector $\bmy_{i}$,
and we use the following shorthand in this section
(note that these differ slightly from the shorthand \cref{eq:btlz_btlM:v_update:batch} used above)
\begin{equation}
\begin{aligned}
    \btlz_{i} &\defequ \bbrz_{i}(\bmF_{t-1},\bmv_{t})
    , &
    \btlM_i &\defequ \bmM_i(\bmF_{t-1},\bmv_{t})
    .
\end{aligned}
\end{equation}
We compute the solutions for $\bmf_j$ in parallel as
\begin{align}
    \bmf_{t,j} = \btlR_j^{-1}\btls_j \quad \forall j\in[d].
    \label{eq:F_update}
\end{align}
 \subsection{Memory and Computational Complexity}
 
 The batch algorithm above involves first accessing all $n = \sum_{\ell=1}^L n_\ell$ data vectors to compute the minorizer parameters $\btlM_{i} \in \bbR^{k \times k}$ and $\btlz_{i} \in \bbR^k$ at a cost of $\mathcal{O}(nk^3 + \sum_{\ell=1}^L  \sumigil |\Omega_{i}|k^2)$ flops per iteration and $\mathcal{O}(n(k^2 + k))$ memory elements. Computing $\btlR_j$ and $\btls_j$ incurs a cost of $\mathcal{O}(\sum_{\ell=1}^L \sumigil |\Omega_{i}|(k^2 + k))$ flops for all $j=1,...,d$, and finally solving for the rows of $\bmF$ costs $\mathcal{O}(dk^3)$ flops per iteration. Updating $\bmv$ requires $ \mathcal{O}(\sum_{\ell=1}^L \sumigil |\Omega_{i}| k^2)$ computations.
 
 Since each complete update depends on all $n$ samples, the batch algorithm must have access to the entire dataset at run-time, either by reading over all the data in multiple passes while accumulating the computed terms used to parameterize the minorizers, or by storing all the data at once, which requires $\mathcal{O}(\sum_{i=1}^n |\Omega_i| + d k^2)$ memory. This requirement, combined with the $\mathcal{O}(n)$ inversions of $k \times k$ matrices in each iteration, significantly limits the practicality of the batch algorithm for massive-scale or high-arrival-rate data as well as in infinite-streaming applications.

\section{Proposed Algorithm: SHASTA-PCA}
\label{s,shasta}
\label{sshasta}

The structure of the log-likelihood in \cref{eq:log_likelihood:incremental}
suggests a natural way to perform incremental (in the finite data setting)
or stochastic (in expectation) updates. 
If each data sample from the $\ell$th group is drawn i.i.d.
from the model in \cref{eq:shasta:generative_model}
, then under uniform random sampling of the data entries,
each $\clL_i$ is an unbiased estimator of $\clL$.
Hence, we leverage the work in \citet{Mairal2013_stochastic_mm},
which proposed a stochastic MM (SMM) technique
for optimizing empirical loss functions
from large-scale or possibly infinite data sets.
For the loss function we consider,
these online algorithms have recursive updates
with a light memory footprint that is independent of the number of samples.
  
In the streaming setting with only a single observation $\bmy_{\Omega_t}$ at each time index $t$,
we do not have access to the full batch minorizer in \cref{eq:em:batch_alg},
but rather only a single $\Psi_t(\bmF,\bmv; \btlF, \btlv)$.
Key to our approach, for each $t$,
our proposed algorithm uses $\Psi_t(\bmF,\bmv; \btlF, \btlv)$
to update two separate approximations
to $\Psi(\bmF, \bmv; \btlF, \btlv)$ parameterized by $\bmF$ and $\bmv$, respectively,
i.e., $\Psib_t^{(F)}(\bmF)$ and $\Psib_t^{(v)}(\bmv)$, in an alternating way.
\tmlrrev{While other optimization approaches are possible---for example,
performing block coordinate maximization of a single joint approximate majorizer---%
our novel approach of alternating between the two separate approximate minorizers reduces memory usage and computational overhead in our setting.}
Given a sequence of non-increasing positive weights
$(w_t)_{t\geq0} \in (0,1)$ and positive scalars $c_v$ and $c_F$,
we first update $\bmv$ (with $\bmF$ fixed at $\bmF_{t-1}$) with one SMM iteration,
then update $\bmF$ (with $\bmv$ fixed at $\bmv_{t}$) with another. Namely, we have the noise variance update:
\begin{align}
    \Psib_{t}^{(v)}(\bmv) &= (1-w_t)\Psib_{t-1}^{(v)}(\bmv)
    + w_t \Psi_t(\bmF_{t-1}, \bmv; \bmF_{t-1}, \bmv_{t-1}), \label{eq:shasta:psib_update} \\
    \bmv_{t} &= (1 - c_v) \bmv_{t-1} + c_v \argmax_{\bmv} \Psib_{t}^{(v)}(\bmv)
    \label{eq:shasta:vupdate},
\end{align}
followed by the factor update:
\begin{align}
    \Psib_t^{(F)}(\bmF) &= (1-w_t)\Psib_{t-1}^{(F)}(\bmF) + w_t \Psi_t(\bmF, \bmv_{t}; \bmF_{t-1}, \bmv_{t})\\
    \bmF_{t} &= (1 - c_F) \bmF_{t-1} + c_F \argmax_{\bmF} \Psib_t^{(F)}(\bmF)
    .
    \label{eq:shasta2:Fupdate_iterate}
\end{align}

The iterate averaging updates in \cref{eq:shasta:vupdate} and \cref{eq:shasta2:Fupdate_iterate}
are important to control the distance between iterates
and have both practical and theoretical significance in SMM algorithms
\citep{Strohmeier2020OnlineNT,lyu2024stochastic}.
Empirically, we found that using constant $c_F$ and $c_v$ worked well,
but other iterate averaging techniques are also possible,
such as those discussed in \citet{Mairal2013_stochastic_mm}.
Other ways to control the iterates include optimizing over a trust region,
as done in \citet{Strohmeier2020OnlineNT}.

Since the iterate and the time index are the same in the streaming setting, i.e., $t = i$,
we now denote both the sample and the SMM iteration by $t$ in the remainder of this section. %
We now derive efficient recursive updates and compare the memory and computational costs to the batch algorithm.

\subsection{Optimizing \texorpdfstring{$\bmv$}{v}
 for fixed \texorpdfstring{$\bmF$}{F}}

Similar to \S\ref{chap:shasta:alg:batch}, we use the following shorthands
\begin{equation}
\begin{aligned}
    \btlz_{t} \defequ \bbrz_{t}(\bmF_{t-1},\bmv_{t-1}), \quad
    \btlM_t \defequ \bmM_t(\bmF_{t-1},\bmv_{t-1})
    .
    \label{eq:btlz_btlM:v_update}
\end{aligned}
\end{equation}
%
Now note that
\begin{equation}
    \Psi_t(\bmF_{t-1}, \bmv; \bmF_{t-1}, \bmv_{t-1})
    =
    C_t
    - |\Omega_t| \ln v_{g_t}
    - \frac{\tlrho_{t}}{v_{g_t}}
    ,
\end{equation}
where $C_t$ does not depend on $\bmv$
and
\begin{align}
\tlrho_{t}
&\defequ
\| \bmy_{\Omega_t} - \bmF_{t-1,\Omega_t} \btlz_t \|_2^2 +
v_{t-1,g_t}
\tr(\bmF_{t-1,\Omega_t}'\bmF_{t-1,\Omega_t} \btlM_{t}  )
.
\end{align}
Recall that $g_t \in [L]$ is the group index of the $t\mathrm{th}$ data vector. Thus, it follows that
\begin{equation}
    \Psib_{t}^{(v)}(\bmv)
    =
    C_t'
    +
    \sum_{\ell=1}^L
    - \brtheta_{t,\ell} \ln v_{\ell}
    - \frac{\brrho_{t,\ell}}{v_{\ell}}
    ,
\end{equation}
where $C_t'$ is a constant that does not depend on $\bmv$,
\begin{align}
    \brtheta_{t,g_t}
    &\defequ
    (1-w_t) \brtheta_{t-1,g_t} + w_t |\Omega_t|
    , \label{eq:theta_gt_update}\\
    \brrho_{t,g_t}
    &\defequ
    (1-w_t) \brrho_{t-1,g_t} + w_t \tlrho_{t} \label{eq:rho_gt_update}
    ,
\end{align}
and for $\ell \neq g_t$
\begin{equation}
\begin{aligned}
    \brtheta_{t,\ell}
    &\defequ
    (1-w_t) \brtheta_{t-1,\ell}
    ,
    &
    \brrho_{t,\ell}
    &\defequ
    (1-w_t) \brrho_{t-1,\ell} \label{eq:theta_rho_ell_neq_gt_update}
    .
\end{aligned}
\end{equation}
The $\ell$th term in the sum is optimized
by $v_\ell = \brrho_{t,\ell} / \brtheta_{t,\ell}$,
so
\begin{equation}
    v_{t,\ell}
    = (1 - c_v) v_{t-1,\ell}
    + c_v
    \frac{\brrho_{t,\ell}}{\brtheta_{t,\ell}} \label{eq:v_iterate_update}
    .
\end{equation}

Here the vectors $\bar{\bmtheta}_{t} \in \mathbb{R}^L$ and $ \bar{\bmrho}_{t} \in \mathbb{R}^L$
aggregate past information to parameterize the approximate minorizer in $\bmv$.
Since there is no past information at $t = 0$, we chose to initialize them with zero vectors.
However, in the initial iterations
where no data vectors have been observed for the $\ell \mathrm{th}$ group,
\cref{eq:v_iterate_update} is undefined,
so a valid argument maximizing \cref{eq:shasta:vupdate} is simply $v_{0,\ell}$,
i.e., the initialized value.



\subsection{Optimizing \texorpdfstring{$\bmF$}{F}
 for fixed \texorpdfstring{$\bmv$}{v}}

We now derive the update for $\bmF$ in \cref{eq:shasta2:Fupdate_iterate}.
Holding $\bmv$ fixed at $\bmv_{t}$, let 
\begin{equation}
\begin{aligned}
    \btlz_{t} &\defequ \bbrz_{t}(\bmF_{t-1},\bmv_{t})
    , &
    \btlM_t &\defequ \bmM_t(\bmF_{t-1},\bmv_{t})
    .
\end{aligned}
\end{equation}
Note that these are redefined from \cref{eq:btlz_btlM:v_update},
where $\bmv_{t-1}$ is replaced with $\bmv_t$ after the variance update.
Maximizing the approximate minorizer $\Psib_t^{(F)}(\bmF)$ with respect to $\bmF$
reduces to maximizing $d$ quadratics in the rows of $\bmF$ for $j \in [d]$:
\begin{align} \label{eq:approx_minorizer:F}
    \Psib_t^{(F)}(\bmF) = \sum_{j=1}^d \bmf_j'\bbrs_{t,j} 
  - \bmf_j' \bbrR_{t,j}\bmf_j
  ,
\end{align}
where for $j \in \Omega_t$
\begin{align} 
\bbrR_{t,j} &= (1-w_t) \bbrR_{t-1,j} + w_t  \bmR_{t,j}, \label{eq:Rbt_update} \\
    \bbrs_{t,j} &= (1-w_t) \bbrs_{t-1,j} + w_t  \bms_{t,j},\label{eq:sbt_update}
\end{align}
where 
\begin{equation}
\begin{aligned}
    \bmR_{t,j} &\defequ \frac{1}{2}\left(\frac{1}{v_{t,g_t}}\btlz_{t} \btlz_{t} ' + \btlM_t\right)
    , &
    \bms_{t,j} &\defequ \frac{1}{v_{t,g_t}} y_{tj} \btlz_{t}
    ,\label{eq:Rti_sti} 
\end{aligned}
\end{equation}
and for $j \notin \Omega_t$
\begin{equation}
\begin{aligned} 
\bbrR_{t,j} &= (1-w_t) \bbrR_{t-1,j}
, &
\bbrs_{t,j} &= (1-w_t) \bbrs_{t-1,j}
. \label{eq:Rbt_sbt_update_missing}
\end{aligned}
\end{equation}
The parameters $(\bbrR_{t,j}, \bbrs_{t,j})$ for $j\in[d]$
of the approximate minorizer $\Psib_t^{(F)}(\bmF)$
aggregate past information from previously observed samples,
permitting our algorithm to stream over an arbitrary amount of data while using a constant amount of memory.

Maximizing the approximate minorizer $\Psib_t^{(F)}(\bmF)$ with respect to each row of $\bmF$ yields
\begin{align} \label{eq:row_f_update}
    \hat{\bmf}_j = \bbrR_{t,j}^{-1} \bbrs_{t,j}, \quad i=1,\hdots,d.
\end{align}
%
Since the problem separates in each row of $\bmF$,
this form permits efficient parallel computations.
Further, because
$\htf_j = \bbrR_{t,j}^{-1} \bbrs_{t,j} = \bbrR_{t-1,j}^{-1} \bbrs_{t-1,j}$
for $j \notin \Omega_t$,
we solve the $k \times k$ linear systems in \cref{eq:row_f_update}
only for the rows indexed by $j \in \Omega_t$.
After obtaining the candidate iterate $\hat{\bmF}$ above,
the final step updates $\bmF_{t}$ by averaging in \cref{eq:shasta2:Fupdate_iterate}.

\subsection{Algorithm and Memory/Computational Complexity}

Together, these alternating updates form the Streaming HeteroscedASTic Algorithm for PCA (SHASTA-PCA),
detailed in \cref{alg:shasta2}.

\IncMargin{1.5em}
\begin{algorithm}
\SetAlCapHSkip{.7em}
    \KwInput{Rank $k$, weights $(w_t) \in (0,1]$,
    parameters $c_F, c_v > 0$, initialization parameter $\delta > 0$.}
    
    \KwData{$[\bmy_{1},\hdots,\bmy_{T}]$, $\bmy_t \in \bbR^{d}$,
    group memberships $g_t \in \{1,2,\dots,L\}$ for all $t$, and sets of observed indices
    $(\Omega_1,\hdots,\Omega_T)$, where $\Omega_t \subseteq \{1,\hdots,d\}$.}

    \KwOutput{$\bmF \in \bbR^{d\times k}$, $\bmv \in \bbR^L_+$.}
    
    Initialize $\bmF_0$ and $\bmv_0$ via random initialization\;
    
    Initialize surrogate parameters $\bbrR_{t,j} = \delta \bmI_k$ for $\delta >0$
    and $\bbrs_{t,j} = \bm0_k$ for $j\in [d]$, $\bar{\bmtheta}_{0} =\bbrrho_0 = \bm0_L$ \;
    
    \For{$t = 1,\hdots,T$}
    {

    Fixing $\bmF$ at $\bmF_{t-1}$,
    \begin{enumerate}
        \item Compute $\bar{\bmtheta}_{t}$ and $\bar{\bmrho}_{t}$
        via \cref{eq:theta_gt_update}-\cref{eq:theta_rho_ell_neq_gt_update}.
        \item Compute $\bmv_t$ from \cref{eq:v_iterate_update}.
    \end{enumerate}

    Fixing $\bmv$ at $\bmv_{t}$,
   	\begin{enumerate}
   	    \item Update $\bbrR_{t,j}$ and $\bbrs_{t,j}$ via \cref{eq:Rbt_update}-\cref{eq:Rbt_sbt_update_missing}.
        \vspace{1mm}
   	    \item Compute $\hat{\bmF}$ via \cref{eq:row_f_update} in parallel.
   	    \item $\bmF_{t} = (1-c_F)\bmF_{t-1} + c_F\hat{\bmF}$.
   	\end{enumerate}
    }
\caption{SHASTA-PCA}
\label{alg:shasta2}
\end{algorithm}
\DecMargin{1.5em}

The primary memory requirement of SHASTA-PCA is storing $d+1$ many $k \times k$ matrices
and $k$-length vectors
for the $\bmF$ surrogate parameters
and two additional $L$-length vectors for the $\bmv$ parameters.
Thus, the dominant memory requirement of SHASTA-PCA
is $\mathcal{O}(d(k^2 + k))$ memory elements
throughout the runtime,
which is independent of the number of data samples.

The primary sources of computational complexity arise from:
i) forming $\btlM_t$ at a cost of $\mathcal{O}(|\Omega_t|k^2 + k^3)$ flops,
ii) computing $\btlz_{t}$ at a cost of $\mathcal{O}(|\Omega_t|k^2)$ flops,
iii) forming $\frac{1}{v_{t,g_t}}\btlz_{t}\btlz_{t}' + \btlM_t$
at a cost of $\mathcal{O}(k^2)$ flops,
iv) computing $\rho_{t,\ell}$
at a cost of $\mathcal{O}(|\Omega_t|k^2 + k^3)$ flops
when using an efficient implementation with matrix-vector multiplications,
and v) updating $\bmF_{t}$
at a cost of
$\mathcal{O}(|\Omega_t|(k^3 + k^2))$ for the multiplications and inverses.
In total, each iteration of SHASTA-PCA incurs $\mathcal{O}(|\Omega_t|(k^3 + k^2))$ flops.

As discussed below,
PETRELS \citep{chi_petrels2013} uses rank-one updates to the pseudo-inverses of the matrices
in \cref{eq:petrels_Rti} to avoid computing a new pseudo-inverse each iteration,
but that approach does not apply in our case
since the updates to $\bbrR_{t,j}$ in \cref{eq:Rbt_update} are not rank-one.
Still, updating $\bmF$ only requires inverting $|\Omega_t|$ many $k \times k$ matrices each iteration,
which remains relatively inexpensive since $k \ll d$ and is often small in practice.
\revise{Note that the complexity appears to be the worst for $|\Omega_t| = d$
with the implementation described above, but in this setting,
since all the surrogate parameters are the same,
one can verify that only a single surrogate parameter $\bbrR_t$ (and its inverse) is necessary.
In reality, the worst-case complexity happens when $|\Omega_t| = d-1$,
i.e., when a single entry per column is missing.}
Identifying an approach with improved computational complexity in the highly-sampled setting
remains an interesting future research direction.

\subsubsection{Convergence}
\label{sss:convergence}

Empirically, we observe convergence to a stationary point as the number of samples grows.
Several factors influence how fast the algorithm converges in practice.
Similar to stochastic gradient descent,
the choices of weights $(w_t)$ and iterate averaging parameters $c_F$ and $c_v$
affect both how fast the algorithm converges and what level of accuracy it achieves.
Using larger weights tends to lead to faster convergence
but only to within a larger, suboptimal local region of an optimum.
Conversely, using smaller weights tends to lead to slower progress
but to a tighter region around an optimum.
The amount of missing data also plays a key role.
A higher percentage of missing entries generally requires more samples
or more passes over the data to converge to an optimum.

Our setting differs in several key ways
from the prior works discussed in \S\ref{shasta:sec:rel}
that establish convergence for SMM algorithms.
First, our minorizers are neither Lipschitz smooth
nor strongly concave
(in fact, the minorizer for $v_\ell$ is nonconcave),
so the theory in \citet{Mairal2013_stochastic_mm} and \citet{mensch2017stochastic}
does not directly apply.
Second, our algorithm maximizes the log-likelihood in two blocks of variables,
and does so in an alternating fashion
with two separate approximate minorizers $\Psib_t^{(F)}(\bmF)$ and $\Psib_t^{(v)}(\bmv)$,
which is distinct from the work in \citet{Strohmeier2020OnlineNT}
that alternates updates over the blocks of a single joint approximate minorizer.
Notably, such as in the case of $\bmv$,
we update an aggregation $\Psib_t^{(v)}(\bmv)$
of the restricted minorizers $\Psi_i(\bmF_{i-1}, \bmv; \bmF_{i-1}, \bmv_{i-1})$:
\begin{align}
&
\Psib_{t}^{(v)}(\bmv)
= (1-w_t)\Psib_{t-1}^{(v)}(\bmv) + w_t \Psi_t(\bmF_{t-1}, \bmv; \bmF_{t-1}, \bmv_{t-1}) \nonumber 
\\&
= \sum_{i=0}^t \left[{w_i \prod_{j=i+1}^t (1-w_j)}\right] \, \Psi_i(\bmF_{i-1}, \bmv; \bmF_{i-1}, \bmv_{i-1}).
\label{eq:aggregate_Psi_i}
\end{align}
The dependence of $\Psib_t^{(v)}(\bmv)$ on all past iterates $\{\bmF_i\}_{i=0}^t$
in the first argument of each $\Psi_i$ in \cref{eq:aggregate_Psi_i}
precludes using the analysis of \citet{Strohmeier2020OnlineNT} for BCD
of a single approximate minorizer each iteration.
\tmlrrev{To our knowledge, no existing work establishes convergence for this setting.}
However, we conjecture that similar convergence guarantees are possible for SHASTA-PCA
since both $\Psib_t^{(F)}(\bmv)$ and $\Psib_t^{(v)}(\bmv)$ retain many of the same MM properties
used to analyze SMM algorithms.
\tmlrrev{Since the convergence results possible for non-convex problems using stochastic optimization are typically weak,
we leave the proof of convergence to future work.}


\subsection{Connection to recursive least squares and HePPCAT}

The SHASTA-PCA 
update for the factors
in \cref{eq:row_f_update} 
resembles recursive least squares (RLS) algorithms like PETRELS \citep{chi_petrels2013}.
The objective function considered in \citet{chi_petrels2013} is, in fact,
the maximization of our complete log-likelihood in the homoscedastic setting
with respect to the factors $\bmF$ and latent variables $\bmz_t$
without the $\ell_2$ penalty on $\bmz_t$ in \cref{eq:em:complete_likelihood_i}.
PETRELS first estimates the minimizer to $\bmz_t$ via the pseudo-inverse solution
and then updates each row $\bmf_j$ by computing \cref{eq:row_f_update}
using similar updates to the minorizer parameters:
\begin{alignat}{3}
    \text{for } j \in \Omega_t &: \quad
    &
    \bbrR_{t,j} &= \lambda \bbrR_{t-1,j} +  \hat{\bmz}_t \hat{\bmz}_t',  \label{eq:petrels_Rti} & \qquad
    \bbrs_{t,j} &= \lambda \bbrs_{t-1,j} +  y_{t,j}  \hat{\bmz}_t
    , \\
    \text{for } j \not\in \Omega_t &:
    &
    \bbrR_{t,j} &= \lambda \bbrR_{t-1,j}
    , &
    \bbrs_{t,j} &= \lambda \bbrs_{t-1,j}
    ,
\end{alignat}
where $\hat{\bmz}_t = \bmF_{t-1,\Omega_t}^\dagger \bmy_{\Omega_t}$ and $\lambda \in (0,1)$
is a forgetting factor that exponentially downweights the importance of past data.
In the stochastic MM framework,
$w_t$ plays an analogous role to $\lambda$
by exponentially down-weighting surrogates constructed from historical data.

However, there are some important differences.
Here, the complete data log-likelihood effectively introduces Tikhonov regularization on $\bmz_t$,
where the Tikhonov regularization parameter is learned by estimating the noise variances.
The PETRELS objective function can similarly incorporate regularization on the weights,
but with a user-specified hyperparameter.
It is well known that the appropriate hyperparameter in Tikhonov regularization
depends on the noise variance of the data \citep{o2001near, cao2020heteroskedastic}.
Here, SHASTA-PCA implicitly learns this hyperparameter
as part of the maximum-likelihood estimation problem
for the unknown heterogeneous noise variances.

PETRELS can also be thought of as a stochastic second-order method
that quadratically majorizes the function in $\bmF$ at each time $t$
using the pseudo-inverse solution of the weights given some estimate $\bmF_{t-1}$.
Our algorithm optimizes a similar quadratic majorizer in $\bmF$ for each $t$.
While the pseudo-inverse solution for maximizing the complete data log-likehood
in \cref{eq:em:complete_likelihood_i} with respect to $\bmz_t$,
or equivalently the conditional mean of $\bmz_t$,
appears in the update of $\bmF$ through $\bbrz_{t}$,
the update additionally leverages the covariance of the latent variable's conditional distribution
and, perhaps most importantly,
an inverse weighting according to the learned noise variances
\tmlrrev{that downweights noisier data samples.}


\revise{SHASTA-PCA also has connections to the HePPCAT algorithm \citep{hong2021heppcat}.
Indeed, the SHASTA-PCA updates of $\bmF$ and $\bmv$ closely resemble---%
and can be interpreted as stochastic approximations to---%
HePPCAT's EM updates of $\bmF$ and $\bmv$ in \citet[eqn.~(8)]{hong2021heppcat}
and \citet[eqn.~(15)]{hong2021heppcat}, respectively.
More precisely, each $\bbrs_{t,j}$ approximates each column of
\begin{align*}
    \sum_{\ell=1}^L \frac{\bbrZ_{t,\ell} \bmY_\ell'}{v_{t,\ell}}
\end{align*}
of \citet[eqn.~(8)]{hong2021heppcat} and each $\bbrR_{t,j}$ approximates the matrix
\begin{align*}
      \sum_{\ell=1}^L
        \frac{\bbrZ_{t,\ell}\bbrZ_{t,\ell}'}{v_{t,\ell}}
        + n_\ell \bmM_{t,\ell}
\end{align*}
in \citet[eqn.~(8)]{hong2021heppcat}.
However, each of these terms in SHASTA-PCA depends on the observed data coordinate
in the update of the corresponding row of $\bmF$.
Since each row of $\bmF$ depends on a different $\bbrR_{t,j}$
for each $j \in [d]$ \tmlrrev{due to missing data},
we cannot use the SVD factorization of $\bmF_t$
to expedite the inverse computation as in HePPCAT \citep[eqn.~(9)]{hong2021heppcat}.
Other minorizers for the $\bmv$ update that were considered in \citet{hong2021heppcat},
such as the difference of concave, quadratic solvable, and cubic solvable minorizers,
may also have possible stochastic implementations.
We leave these possible approaches to future work since they did not appear to result in more efficient updates.
}

\section{Results} \label{sec:results}

\subsection{Incremental computation with static subspace}
\label{shasta:subsection:incremental-full}
This section considers the task of estimating a static planted subspace from low-rank data corrupted by heterogeneous noise. We generate data according to the model in \cref{eq:shasta:generative_model} with ambient dimension $d=100$ from a rank-3 subspace with squared singular values $[4,2,1]$,
drawing 500 samples with noise variance $10^{-2}$, and 2,000 samples with noise variance $10^{-1}$.
We draw an orthonormal subspace basis matrix $\bmU \in \bbR^{100 \times 3}$ uniformly at random from the Stiefel manifold, 
and set the planted factor matrix to be $\bmF = \bmU \sqrt{\bmlambda}$.

After randomly permuting the order of the data vectors,
we compared SHASTA-PCA to PETRELS
and the streaming PCA algorithm GROUSE \citep{grouse} (which has recently been shown to be equivalent to Oja's method \citep{Oja1982SimplifiedNM} in \citet{balzano2022equivalence})
that estimates a subspace from rank-one gradient steps on the Grassmann manifold,
with a tuned step size of $0.01$.\footnote{All experiments were performed in Julia on a 2021 Macbook Pro with the Apple M1 Pro processor and 16 GB of memory. We reproduced and implemented all algorithms ourselves from their original source works.}
SHASTA-PCA jointly learns both the factors $\bmF$ and noise variances $\bmv$ from each streaming observation. For SHASTA-PCA, we used $w_t = 1/t$ (where $t$ is the time index), $c_F = c_v = 0.1$ and initialize the parameters $\bbrR_t(i) = \delta \bmI$ with $\delta = 0.1$ for both SHASTA-PCA and PETRELS.
We initialized each streaming algorithm with the same random $\bmF_0$,
and each entry of $\bmv_0$ for SHASTA-PCA uniformly at random between 0 and 1.
We set the forgetting parameter in PETRELS to $\lambda = 1$, corresponding to the algorithm's batch mode.
As a baseline, we compared to batch algorithms for fully-observed data: HePPCAT \citep{hong2021heppcat} with 100 iterations, which we found to be sufficient for convergence, and homoscedastic probabilistic PCA (PPCA) \citep{tipping1999ppc} on the full data. In addition, we computed PPCA over each data group individually, denoted by ``G1'' (``G2'') in the legend of \cref{fig:batch:100p:static} corresponding to group 1 (2) with 500 (2,000) samples with noise variances $10^{-2}$ ($10^{-1}$) respectively.

The first experiment in \cref{fig:batch:100p:static} compares each algorithm in the fully observed data setting,
where the streaming algorithms compute $\bmF$ and $\bmv$ incrementally using a single vector in each iteration.
Given the planted model parameters $\bmF^*$ and $\bmv^*$ and their log-likelihood value $\clL^*:=\clL(\bmF^*,\bmv^*)$,
the left plot in \cref{fig:batch:100p:static} shows the normalized log-likelihood $\clL(\bmF_t,\bmv_t) - \clL^*$
with respect to the full dataset in \cref{eq:log_likelihood:incremental}
for each iteration of SHASTA-PCA compared to the batch algorithm baselines.
Because GROUSE and PETRELS do not estimate the noise variances, we omit them from this plot.
The right plot in \cref{fig:batch:100p:static} shows convergence of the $\bmF$ iterates
with respect to the normalized subspace error $\frac{1}{k}\|\hat{\bmU}_t \hat{\bmU}_t' - \bmU \bmU'\|_F^2$
for the estimate $\hat{\bmU}_t \in \bbR^{d \times k}$ of the planted subspace $\bmU$;
for SHASTA-PCA and PETRELS, we compute $\hat{\bmU}_t$ by taking the $k$ left singular vectors of $\bmF$.
Each figure plots the mean of 50 random initializations in bold dashed traces, where their standard deviations are displayed as ribbons.
The experiment in \cref{fig:batch:50p:static} then subsamples 50\% of the data entries uniformly at random,
inserting zeros for the missing entries for methods that require fully sampled data,
and compares the same statistics across the algorithms.

As expected, when the samples were fully observed, PETRELS converged to the same log-likelihood and subspace error for each set of training data 
as the batch algorithms that assume homoscedastic noise,
and SHASTA-PCA converged to the same log-likelihood value and subspace error as HePPCAT.
Here we see the advantage of using heteroscedastic data analysis.
Instead of discarding the samples from either data group or combining them in a single PPCA,
the heteroscedastc PPCA algorithms leverage both the ``clean'' samples, the additional ``noisy'' samples,
and the noise variance estimates to produce better subspace estimates.
With many missing entries (imputed with zeros),
the batch algorithms' subspace estimates quickly deteriorated,
as seen on the right-hand side of \cref{fig:batch:50p:static}.
Out of the streaming PCA algorithms for missing data,
SHASTA-PCA again attained the best subspace estimate compared to GROUSE and PETRELS.
\oneortwocol{
\begin{figure*}
    \centering
    \subfloat[Fully observed data.]{\includegraphics[trim={1cm 0 5.12cm 0},clip,height = 1.75in]{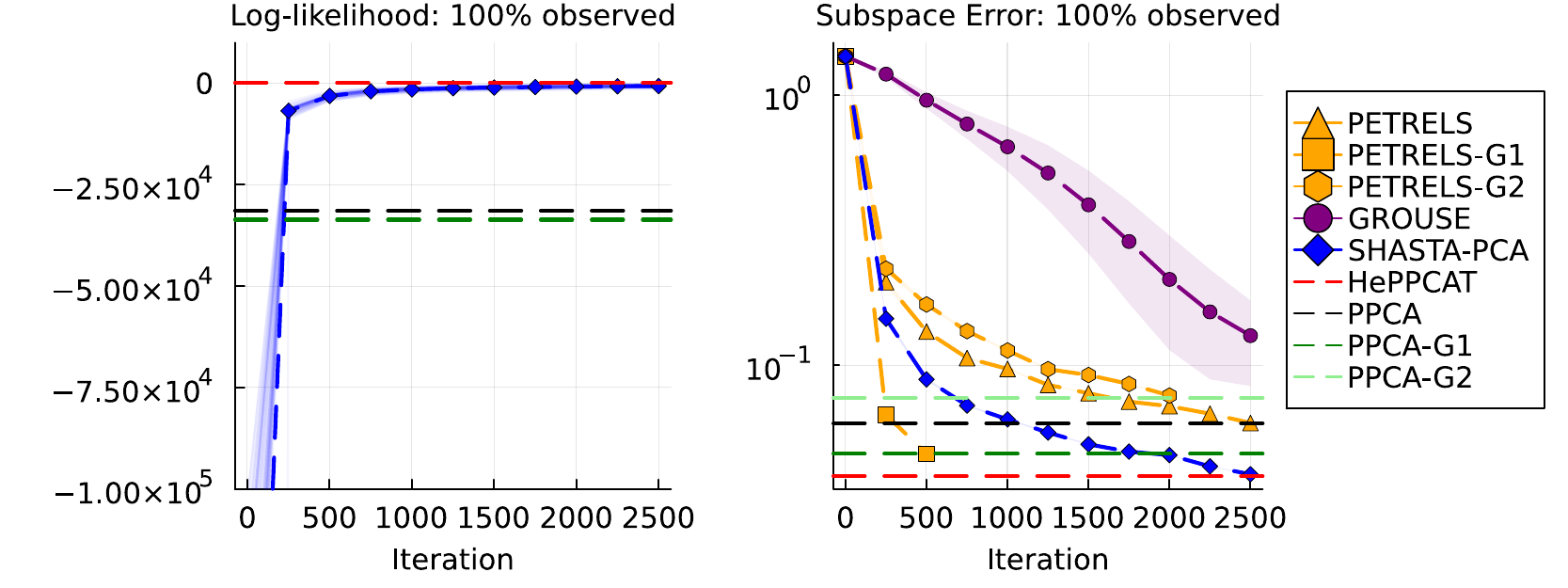}
    \label{fig:batch:100p:static}}
    \subfloat[Data with 50\% entries observed uniformly at random.]{\includegraphics[trim={1cm 0 0cm 0},clip,height = 1.75in]{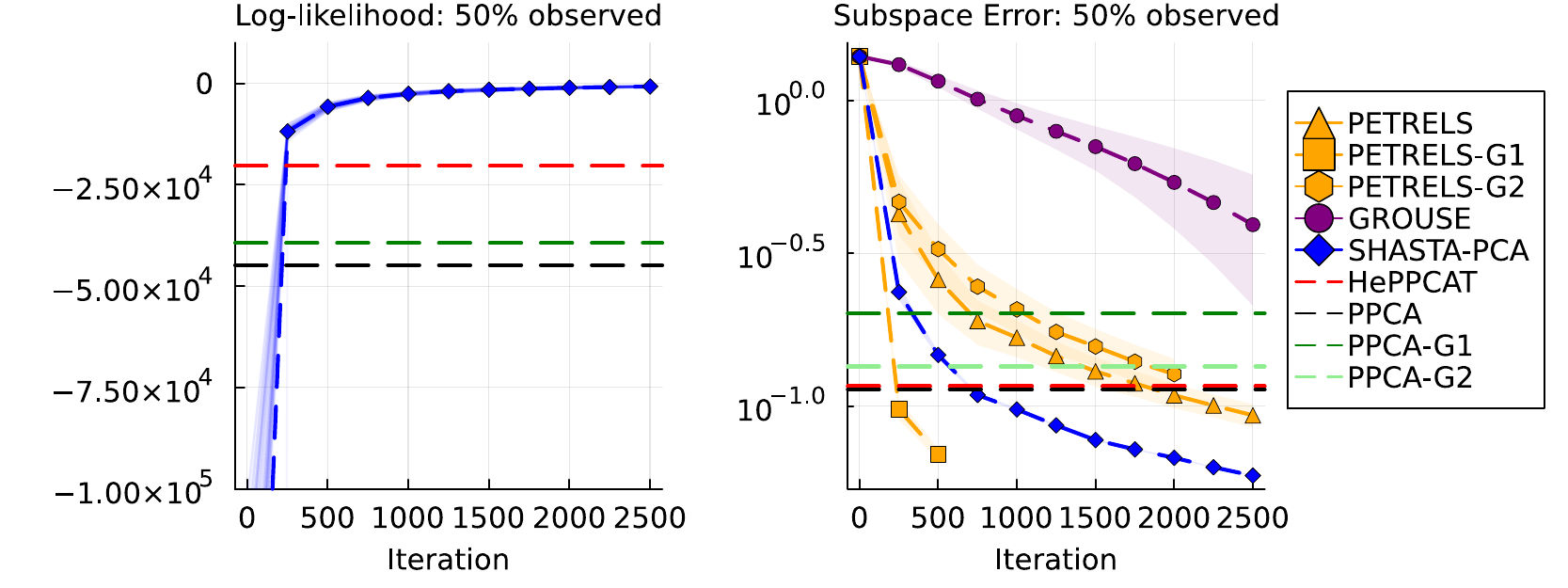}
    \label{fig:batch:50p:static}}
    
    \caption{Incremental computation (with one pass) over batch data generated from a static subspace for $d = 100$, $n_1 = 500$, $n_2 = 2{,}000$, $v_1 = 10^{-2}$, and $v_2 = 10^{-1}$. Horizontal dashed lines correspond to the terminal values of batch algorithms, and the horizontal axis refers to streaming iteration for online algorithms.}
\end{figure*}
}{
\begin{figure}
    \centering
    \subfloat[Fully observed data.]{\includegraphics[height = 1.65in]{figs/synthetic-october-2022/online_algs_iteration-d_100-n_500,2000-observedPercent_100-v1_1.00E-02-v2_1.00E-01_version2.pdf}
    \label{fig:batch:100p:static}}
    \\
    \subfloat[Data with 50\% entries observed uniformly at random.]{\includegraphics[trim={0 0 0cm 0},clip,height = 1.65in]{figs/synthetic-october-2022/online_algs_iteration-d_100-n_500,2000-observedPercent_50-v1_1.00E-02-v2_1.00E-01_version2.pdf}
    \label{fig:batch:50p:static}}
    
    \caption{Incremental computation (with one pass) over batch data generated from a static subspace for $d = 100$, $n_1 = 500$, $n_2 = 2{,}000$, $v_1 = 10^{-2}$, and $v_2 = 10^{-1}$. Horizontal dashed lines show the terminal values for the batch algorithms, and the horizontal axis shows the iteration index for the online algorithms.}
\end{figure}
}

\subsection{Dynamic subspace}

\begin{figure}
    \centering
    {\includegraphics[height=2in]{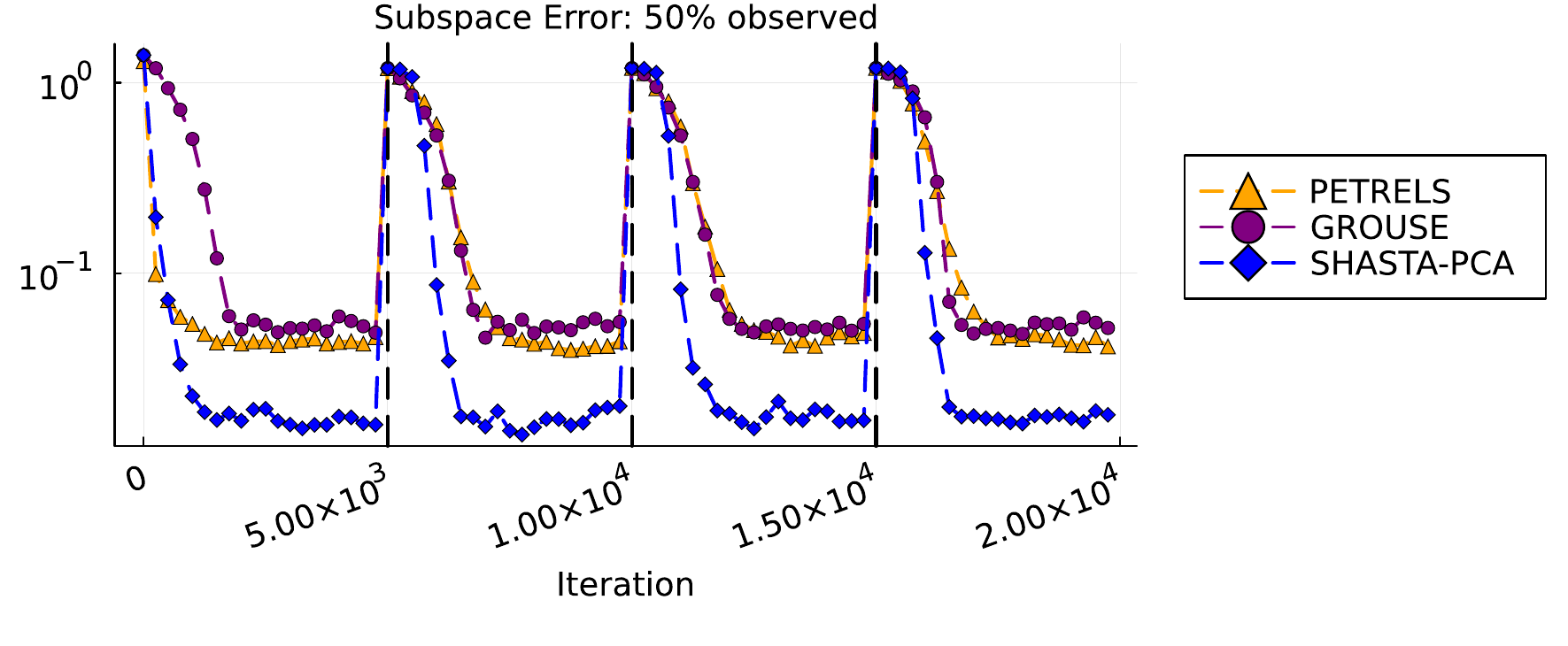}
    }
    \caption{Dynamic tracking of rapidly shifting subspace with 50\% of the entries observed uniformly at random using SHASTA-PCA versus streaming PCA algorithms that assume homoscedastic noise. Here, $d=100$ and 20\% of the data has noise variance $10^{-4}$ and $80\%$ of the data has noise variance $10^{-2}$. Iterations refers to the number of streamed data vectors.}
    \label{fig:exp:synth:dynamic}
\end{figure}

This section studies how well SHASTA-PCA can track a time-varying subspace. We generate 20,000 streaming data samples according to the model in \cref{eq:shasta:generative_model} for $L=2$ groups with noise variances $v_1 = 10^{-4}$ and $v_2 = 10^{-2}$. We use a randomly drawn $\bmF = \bmU \sqrt{\bmlambda}$, where $d=100$ and $\bmlambda = [4,2,1]$. The data samples are drawn from the two groups with 20\% and 80\% probability, respectively. We then observe 50\% of the entries selected uniformly at random. To simulate dynamic jumps of the model, we set the planted subspace $\bmU$ to a new random draw every 5000 samples and compare the subspace errors of the various methods with respect to the current $\bmU$ over time. Here, we use the parameters $w_t = 0.01, c_F = 0.01$, and $c_v = 0.1$ for SHASTA-PCA. After hyperparameter tuning, we set the step size of GROUSE to be 0.02, and we set $\lambda = 0.998$ for PETRELS. Each algorithm is initialized with the same random factors $\bmF_0$, and SHASTA-PCA's noise variances are initialized uniformly at random between 0 and 1. \cref{fig:exp:synth:dynamic} shows SHASTA-PCA outperforms the streaming PCA algorithms that assume homoscedastic noise by half an order of magnitude. The results highlight how the largest noise variance dominates the streaming PCA algorithms' subspace tracking performance while SHASTA-PCA obtains more faithful estimates by accounting for the heterogeneity.

\subsection{Dynamic noise variances}

In some applications, due to temperature, age, or change in calibration, the quality of the sensor measurements may also change with time \citep{jun2003drift}, thereby affecting the levels of noise in the data. To study the performance of SHASTA-PCA in these settings, we generate samples from the planted model described above where we change the noise variances over time while keeping the subspace stationary. As before, SHASTA-PCA is initialized at a random $(\bmF_0, \bmv_0)$. \cref{fig:exp:dynamic_v_plots:v1:verr,fig:exp:dynamic_v_plots:v1:Uerr} show the estimated noise variances and the subspace error as we double the noise variance of the first group every 5,000 samples. \cref{fig:exp:dynamic_v_plots:v2:verr,fig:exp:dynamic_v_plots:v2:Uerr} repeat the experiment but double the noise variance of the second group instead. As $v_1$ increases and the cleaner group becomes noisier, the data becomes noisier overall and also closer to homoscedastic. SHASTA-PCA's estimate of the subspace degrades and approaches the estimates obtained by PETRELS and GROUSE. On the other hand, as the noisier group gets even noisier, the quality of the PETRELS subspace estimate deteriorates in time whereas SHASTA-PCA remains robust to the added noise by leveraging the cleaner data group. In both instances, GROUSE appears to oscillate about an optimum in a region whose size depends on the two noise variances. 
The variance estimates demonstrate how SHASTA-PCA can quickly adapt to changes in the noise variances; SHASTA-PCA adapted here within less than 1,000 samples.

\oneortwocol{
\begin{figure}
\centering
\label{fig:exp:dynamic_v_plots}
\subfloat[Estimated variances for varying $v_1$.]{\includegraphics[trim={0 0 0cm 0},clip, width=1.65in]{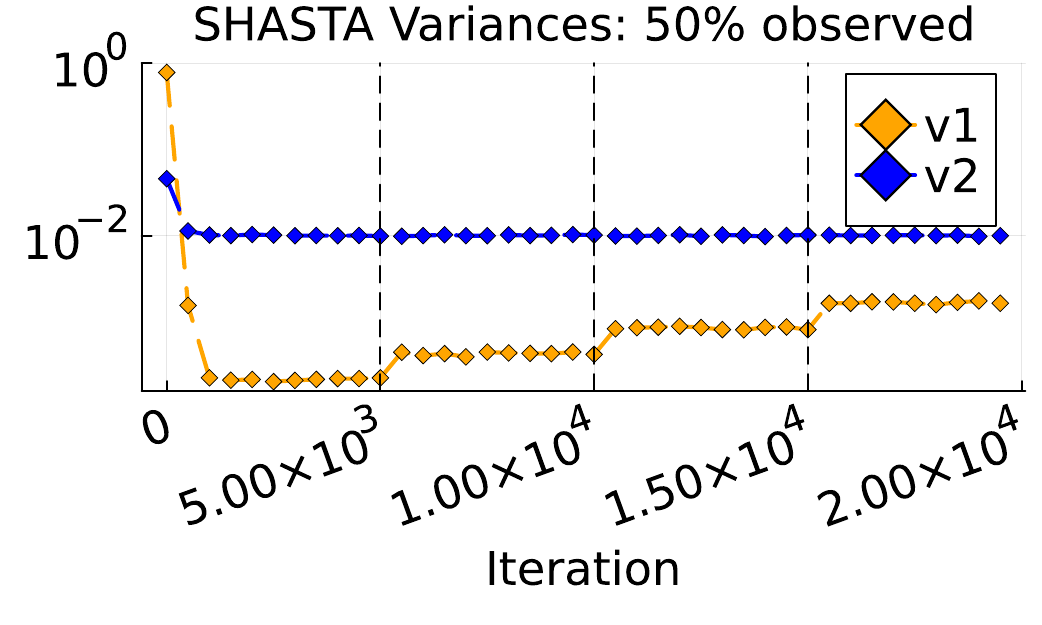}
\label{fig:exp:dynamic_v_plots:v1:verr}
}
\hspace{1cm}
\subfloat[Estimated variances for varying $v_2$]{\includegraphics[trim={0 0 0cm 0},clip,width=1.65in]{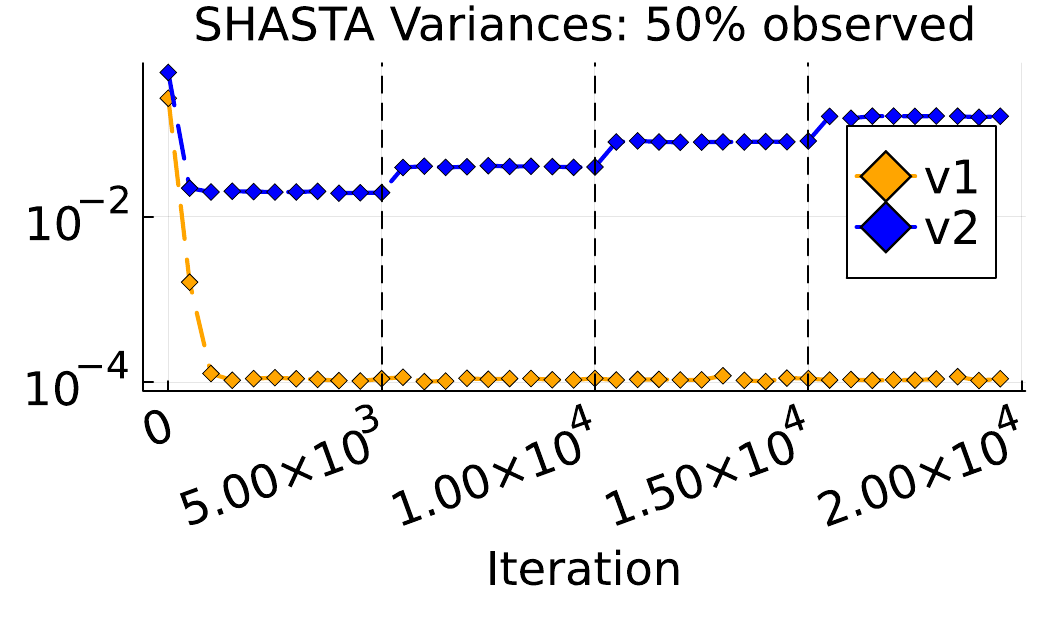}
\label{fig:exp:dynamic_v_plots:v2:verr}}\\

\subfloat[Subspace error with varying $v_1$.]{\includegraphics[clip,width=1.65in]{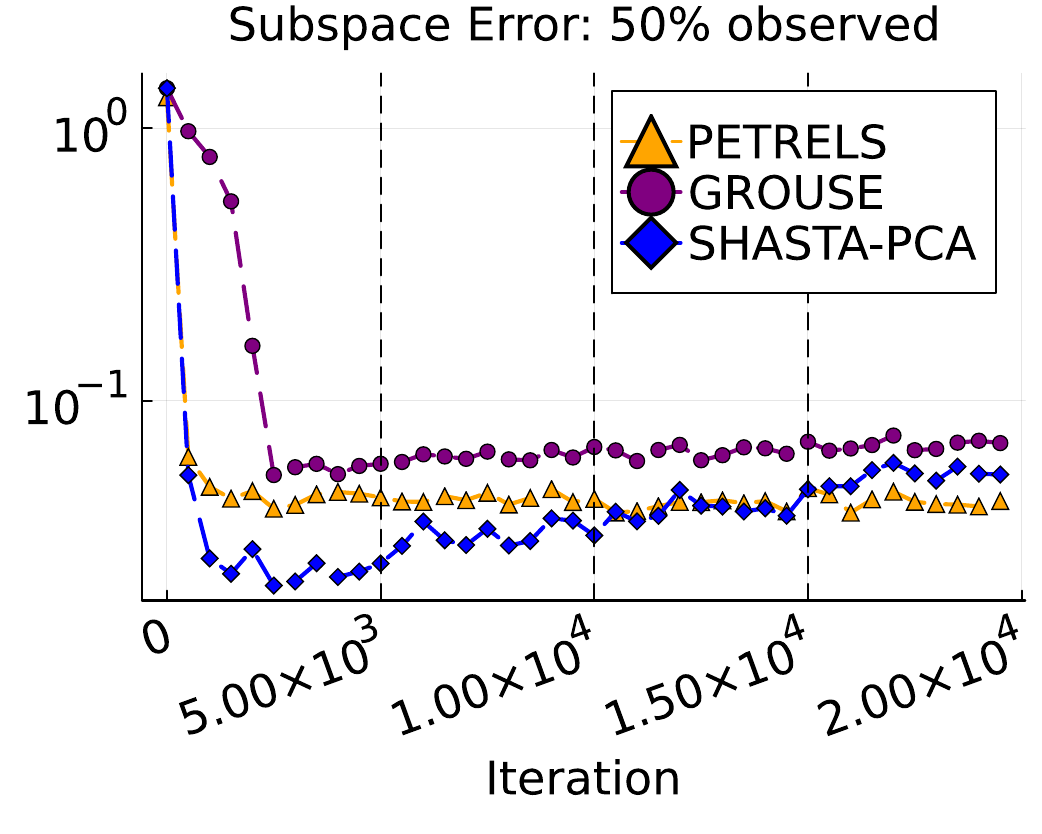}
\label{fig:exp:dynamic_v_plots:v1:Uerr}}
\hspace{1cm}
\subfloat[Subspace error with varying $v_2$.]{\includegraphics[width=1.65in]{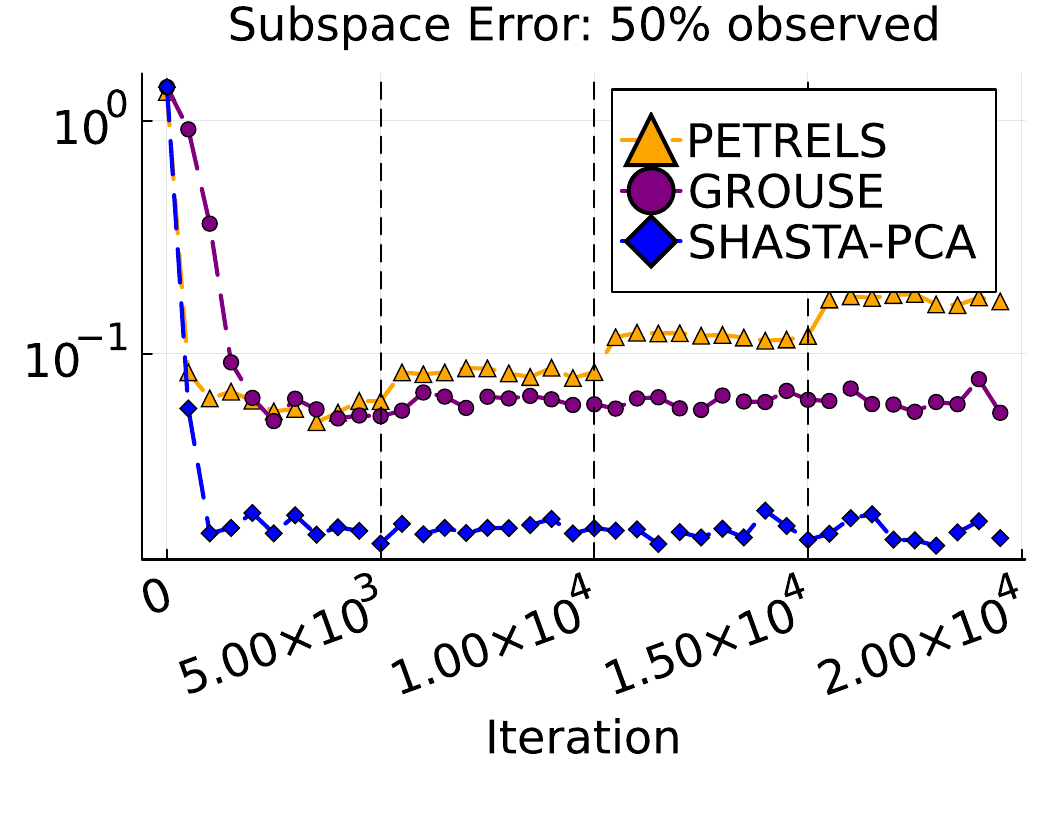}
\label{fig:exp:dynamic_v_plots:v2:Uerr}}
\caption{Experiments with changing variances across time (iterations) for a single $\bmF \in \mathbb{R}^{100 \times 3}$ starting with planted noise variances $\bmv = [10^{-4}, 10^{-2}]$ with 50\% of the entries observed uniformly at random. The vertical dashed lines indicate points at which we double one of the planted variances. The top plots show the estimated variances from SHASTA-PCA, and the bottom plots show the subspace error.}
\end{figure}
}{
\begin{figure}
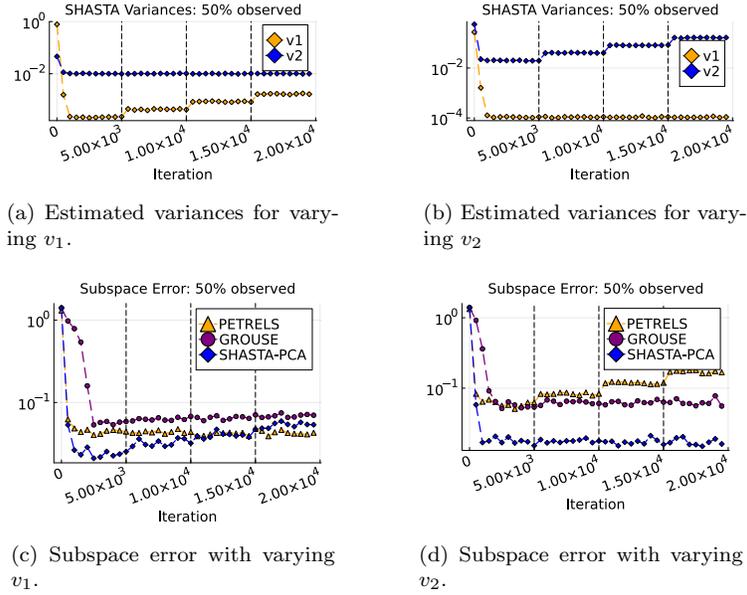

\centering
\label{fig:exp:dynamic_v_plots}
\subfloat[Estimated variances for varying $v_1$.]{\includegraphics[trim={0 0 0cm 0},clip, width=2.5in]{figs/synthetic-december-2022/online_algs_dynamic_v1-v1_error-d_100-n_20000-observedPercent_50-v1_1.60E-03-v2_1.00E-02_v2.pdf}
\label{fig:exp:dynamic_v_plots:v1:verr}
}
\hspace{2mm}
\subfloat[Estimated variances for varying $v_2$]{\includegraphics[trim={0 0 0cm 0},clip,width=2.5in]{figs/synthetic-december-2022/online_algs_dynamic_v2-v_error-d_100-n_20000-observedPercent_50-v1_1.00E-04-v2_1.60E-01_v2.pdf}
\label{fig:exp:dynamic_v_plots:v2:verr}}\\

\subfloat[Subspace error with varying $v_1$.]{\includegraphics[clip,width=2.5in]{figs/synthetic-december-2022/online_algs_dynamic_v1-U_error-d_100-n_20000-observedPercent_50-v1_1.60E-03-v2_1.00E-02_v2.pdf}
\label{fig:exp:dynamic_v_plots:v1:Uerr}}
\hspace{2mm}
\subfloat[Subspace error with varying $v_2$.]{\includegraphics[width=2.5in]{figs/synthetic-december-2022/online_algs_dynamic_v2-U_error-d_100-n_20000-observedPercent_50-v1_1.00E-04-v2_1.60E-01_v2.pdf}
\label{fig:exp:dynamic_v_plots:v2:Uerr}}
\caption{Experiments with changing variances across time (iterations) for a single $\bmF \in \mathbb{R}^{100 \times 3}$ starting with planted noise variances $\bmv = [10^{-4}, 10^{-2}]$ with 50\% of the entries observed uniformly at random. The vertical dashed lines indicate points at which we double one of the planted variances. The top plots show the estimated variances from SHASTA-PCA, and the bottom plots show the subspace error.}
\end{figure}
}

\subsection{Computational timing experiments}

\begin{figure}
\centering
\subfloat[]{\includegraphics[trim={15 0 0cm 0},clip,height=1.75in]{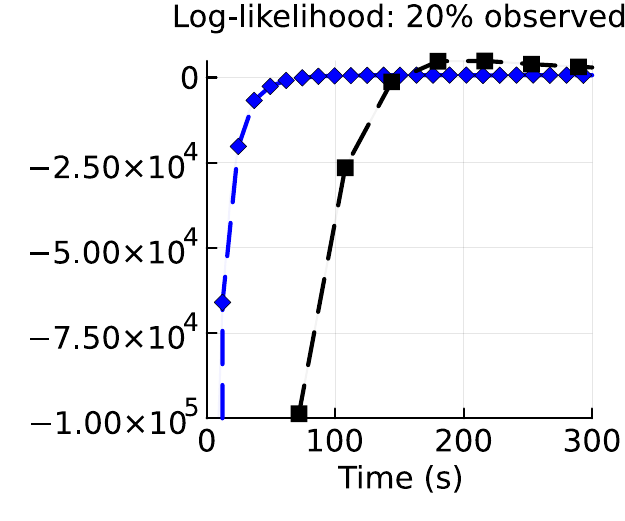}}
\subfloat[]{\includegraphics[height=1.75in]{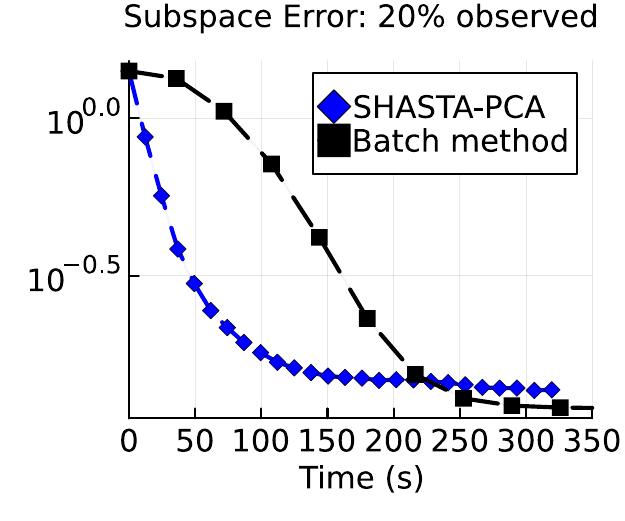}}

\caption{Log-likelihood values and subspace errors versus elapsed wall clock time for one run of SHASTA-PCA versus the batch method in \S\ref{chap:shasta:alg:batch} on 2GB of synthetic data: $d=1{,}000$, $k=3$, $\bmn = [50{,}000, ~ 200{,}000]$, and $\bmv = [0.1, 1]$. Both algorithms used the same random initialization. Markers for SHASTA-PCA are for every $10{,}000$ vector samples, and markers for the batch method are for each algorithm iteration.}
\label{fig:exp:computation_timing}
\end{figure}

Computational and/or storage considerations can inhibit the use of batch algorithms for large datasets, especially on resource constrained devices. To demonstrate the benefit of SHASTA-PCA in such settings, we generated a 2GB dataset according to our model, where $d=1{,}000$, $\bmlambda = [4,2,1]$, $\bmn = [50{,}000, ~ 200{,}000]$, $\bmv = [0.1, 1]$, and we observed only 20\% of the entries uniformly at random.  For this experiment, we set $w_t = 0.01 / \sqrt{t}$, $c_F = 0.01$, and $c_v = 0.1$ for SHASTA's hyperparameters, and passed over the entire data once. \Cref{fig:exp:computation_timing} compares the convergence in log-likelihood values and subspace errors by elapsed wallclock time for SHASTA and the batch algorithm in \S\ref{chap:shasta:alg:batch}, where both algorithms are randomly initialized from the same random starting iterate $(\bmF_0, \bmv_0)$. SHASTA-PCA rapidly obtained a good estimate of the model, using only roughly 60\% of the time that it took the batch method, all while using only 0.0048\% of the memory per iteration.

\subsection{Real data from astronomy}

We illustrate SHASTA-PCA on real astronomy data from the Sloan Digital Sky Survey (SDSS) Data Release 16 \citep{ahumada202016th} using the associated DR16Q
quasar catalog \citep{lyke2020sloan}. In particular, we considered the subset that was considered in \citet[Section 8]{hong2023owp}; see \citet[Supplementary Material SM5]{hong2023owp} for details about the subset selected and the preprocessing performed. The dataset contains $n = $ 10,459 quasar spectra, where each spectrum is a vector of $d=281$ flux measurements across wavelengths and the data come with associated noise variances.

\begin{figure*}

\centering
\subfloat[]{\includegraphics[width=3in]{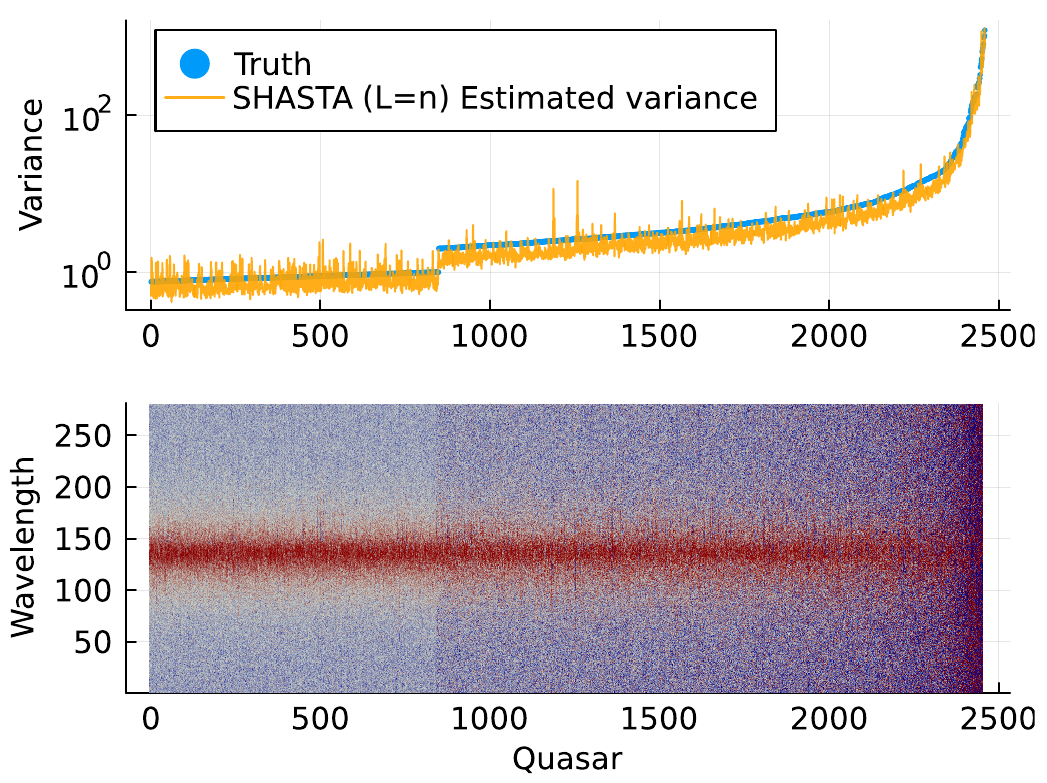} \label{fig:quasar:shastal1_variances_rand_order}
}
\subfloat[]{\includegraphics[width=3in]{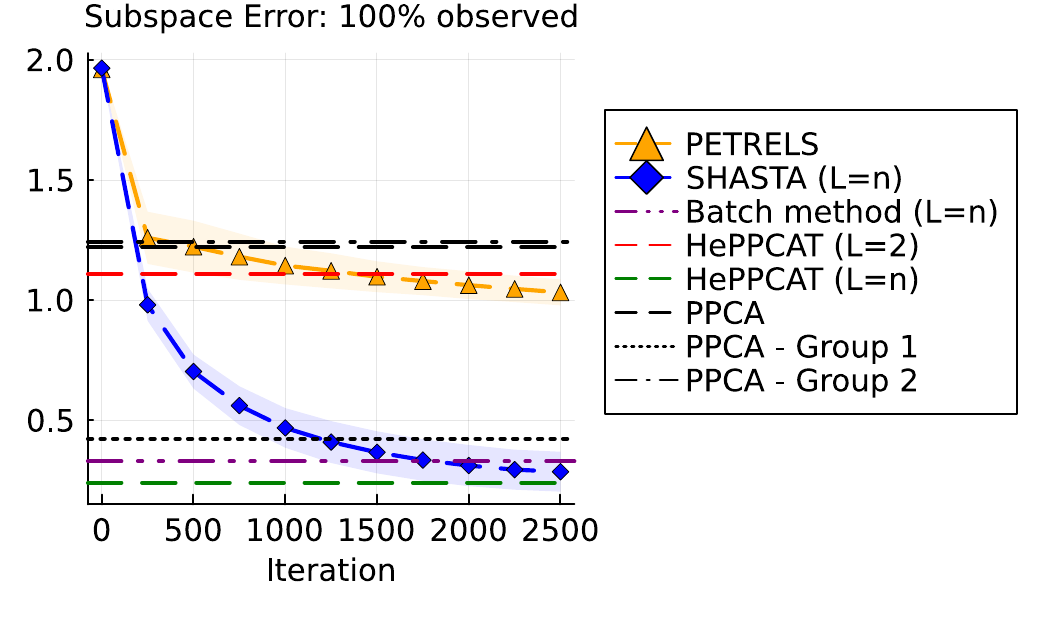}
\label{fig:quasar:shastal1_subspace_error_rand_order}
}

\caption{(a) Visualization of quasar data with true associated variances of each sample plotted by quasar. The estimated variances from a SHASTA-PCA $L=n$ model streaming over the samples (fully sampled, in randomized order) closely matched the true variances. (b) For fully sampled data, the subspace error of SHASTA-PCA converged to the error of HePPCAT's estimated subspace for a $L=n$ model. We repeated the experiment 50 times, each with a different random initialization and order of the samples.}

\label{fig:quasar:shastal1}
\end{figure*}

Ordering the samples from smallest to largest noise variance estimates, we obtained a ``ground-truth'' signal subspace by taking the left $k=5$ singular vectors of the data matrix for the first 2,000 samples with the smallest noise variance estimates. We then formed a training dataset with two groups: first, we collected samples starting from sample index 6,500 to the last index where the noise variance estimate is less than or equal to 1 (7,347); second, we collected training data beginning at the first index where the noise variance estimate is greater than or equal to 2 (8,839) up to the sample index 10,449, excluding the last 10 samples that are grossly corrupted. The resulting training dataset had $n_1 = 848$ and $n_2 = $ 1,611 samples for the two groups, respectively, and had strong noise heteroscedasticity across the samples, where the second group was much noisier than the first. \cref{fig:quasar:shastal1_variances_rand_order} shows the training dataset and the associated noise variance estimates for each sample.

Although we formed the data by combining two groups of consecutive samples, the data actually contained $L=n$ groups since each spectrum has its own noise variance.
This setting allows for a heuristic computational simplification in SHASTA-PCA. Namely, we adapt the $L=1$ model for a single variance $v$ and use separate weights for the SMM updates of $\bmF$ and $v$, where $w_t^{(F)} = 0.001$, $c_F = 0.1$, and $w_t^{(v)} = c_v = 1$ $\forall t$.
The variance update is then equivalent to maximizing $\Psi_t(\bmF_{t-1}, v; \bmF_{t-1}, v_{t-1})$, i.e., the minorizer centered at the previous variance estimate with no memory of previous minorizers. The number of variance EM updates per data vector may be increased beyond just a single update, but in practice, we observed little additional benefit. \Cref{fig:quasar:shastal1} shows how SHASTA-PCA adaptively learned the unknown variances for each new sample and converged to the same level of subspace error as the batch HePPCAT $L=n$ model.

\begin{figure}
    \centering
    \includegraphics[width=4in]{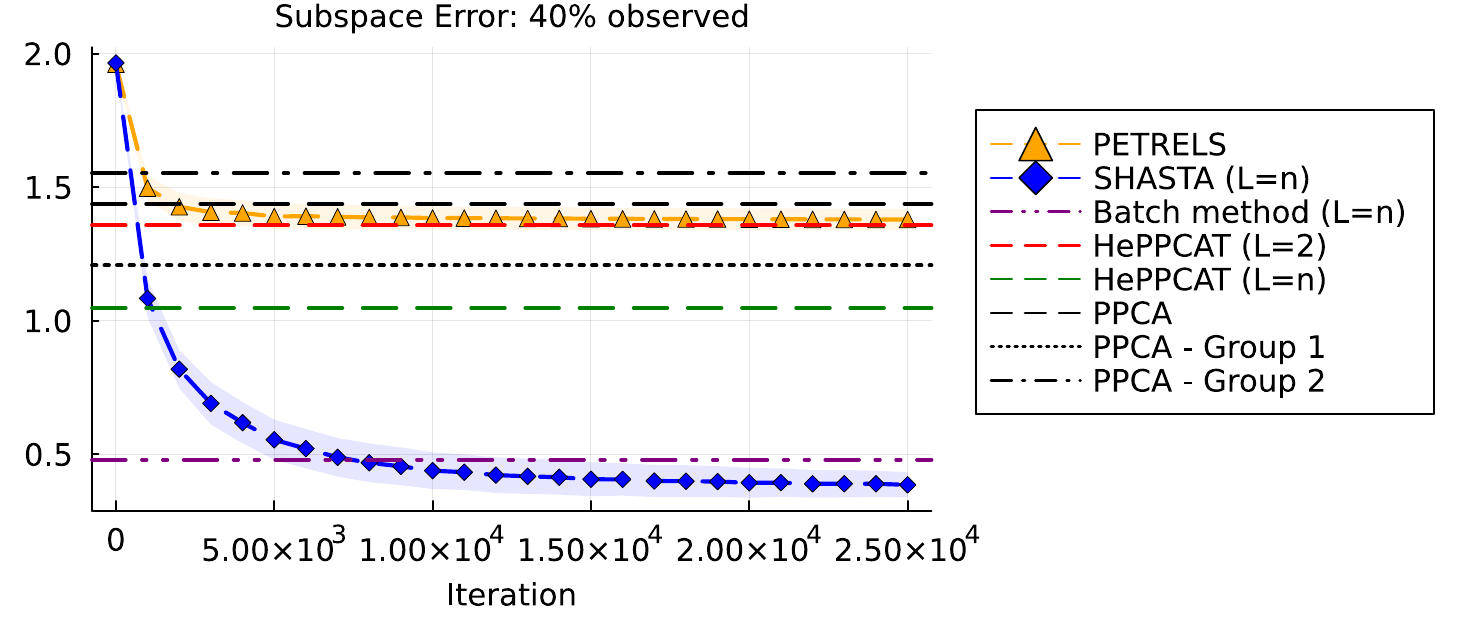}
    \caption{Subspace error for quasar data with 40\% of the entries observed uniformly at random. SHASTA-PCA streams over the full dataset 10 times, with the sample entries missing, where the order of the samples was randomized on each pass. Each experiment was initialized with factors chosen uniformly at random.}
    \label{fig:quasar:shastal1:40p_observed}
\end{figure}

In many modern large datasets, entries may be missing in significant quantities due to sensor failure or time and memory constraints that preclude acquiring complete measurements. Indeed, the experiment designer may only wish to measure a ``sketch'' of the full data to save time and resources and learn the underlying signal subspace from limited observations using an algorithm like SHASTA-PCA. To study this case, we randomly obscured 60\% of the entries uniformly at random and performed 10 passes over the data, randomizing the order of the samples each time. We used the same choice of weights described above to estimate a single variance for every new sample. We initialized with a random $\bmF_0$ and $v_0$ using the zero-padded data. As \cref{fig:quasar:shastal1:40p_observed} shows, SHASTA-PCA had better subspace estimates in this limited sampling setting than the state-of-the-art baseline methods with zero-filled missing entries and/or homoscedastic noise assumptions. Interestingly, the SHASTA-PCA subspace estimate was even better than the batch method in \S\ref{chap:shasta:alg:batch} for the $L=n$ model.

\section{Conclusion \& Future Work}

This paper proposes a new streaming PCA algorithm
(SHASTA-PCA) that is robust to \emph{both} missing data and heteroscedastic noise across samples.
SHASTA-PCA only requires a modest amount of memory
that is independent of the number of samples and has efficient updates that can scale to large datasets.
The results showed significant improvements over state-of-the-art streaming PCA algorithms
in tracking nonstationary subspaces under heteroscedastic noise
and significant improvement over a batch algorithm in speed.

There are many future directions building on this work.
An interesting line of future work is to establish convergence guarantees for SHASTA-PCA,
\tmlrrev{particularly since our optimization approach is unique among other works using stochastic MM.}
Second, while each update of a row in $\bmF$ is relatively cheap,
it still requires inverting $|\Omega_t|$ many $k \times k$ matrices per data vector,
which is particularly wasteful when $|\Omega_t| = d$
since this is equivalent to the fully-sampled log-likelihood,
which only requires inverting a single $k \times k$ matrix to update $\bmF$.
It may be possible to find other surrogate functions
that would avoid this large number of small inverses in each iteration.
Finally, although SHASTA-PCA enjoys lightweight computations in each iteration,
achieving rapid convergence can depend on carefully tuning the weights $w_t$
and the parameters $c_F$ and $c_v$.
Adaptively selecting these parameters with stochastic MM techniques
and developing theory to guide the selection of these parameters remain open problems.

\bibliography{refs_shasta,refs_dahong}
\bibliographystyle{tmlr}

\end{document}